\documentclass[preprintnumbers,
amsmath,amssymb,floatfix,10pt,prd,onecolumn,
superscriptaddress,longbibliography,nofootinbib]{revtex4-2}
\usepackage{bm}
\usepackage{amsfonts}
\usepackage{latexsym}
\usepackage{amsmath}
\usepackage{textcomp}
\usepackage{float}
\usepackage{booktabs}
\usepackage{dcolumn}
\usepackage{dsfont}
\usepackage{ragged2e}
\usepackage{epsfig}
\usepackage[dvipsnames]{xcolor}
\usepackage{hyperref}
\hypersetup{           
	colorlinks=true,                
	breaklinks=true,                
	urlcolor= OrangeRed,                
	linkcolor= magenta,                
	bookmarksopen=false,
	filecolor=black,
	citecolor=red,
	linkbordercolor=blue}
\usepackage{graphicx}
\usepackage{orcidlink}
\usepackage[T1]{fontenc}

\begin{document}
	\title{Fractional Holographic Dark Energy Driven Reconstruction of $f(Q)$ Gravity and its Cosmological Implications}
	
   \author{Rajdeep Mazumdar \orcidlink{0009-0003-7732-875X}}
	\email[Corresponding author: ]{rajdeepmazumdar377@gmail.com}
	\affiliation{%
		Department of Physics, Dibrugarh University, Dibrugarh, Assam, India, 786004}

  \author{Kalyan Malakar\orcidlink{0009-0002-5134-1553}}%
\email{kalyanmalakar349@gmail.com}
\affiliation{Department of Physics, Dibrugarh University, Dibrugarh, Assam, India, 786004}
\affiliation{Department of Physics, Silapathar College, Dhemaji, Assam, India, 787059}

	
	\author{Kalyan Bhuyan\orcidlink{0000-0002-8896-7691}}%
	\email{kalyanbhuyan@dibru.ac.in}
	\affiliation{%
		Department of Physics, Dibrugarh University, Dibrugarh, Assam, India, 786004}%
	\affiliation{Theoretical Physics Divison, Centre for Atmospheric Studies, Dibrugarh University, Dibrugarh, Assam, India 786004}

	
\begin{abstract}
In order to explain the late-time acceleration of the Universe, we present a reconstructed version of the $f(Q)$ gravity theory in this work, which is inspired by the integrating the fractional holographic dark energy with the Hubble horizon as the infrared cutoff. This reconstructed $f(Q)$ gravity model shows a geometrically motivated dark energy component and naturally recovers General Relativity in the appropriate limit. The free parameters of the model are constrained using the latest DESI BAO data, previous BAO compilations (P-BAO), and cosmic chronometer (CC) datasets through a Markov Chain Monte Carlo (MCMC) analysis. The reconstructed Hubble parameter $H(z)$ exhibits excellent consistency with observational data, with high values of $R^2$ and low values of $\chi^2_{\min}$, AIC, and BIC, confirming the model's strong statistical performance relative to $\Lambda$CDM. With current $q(0) \in [-0.40, -0.32]$ and a transition redshift $z_{\text{tr}} \sim 0.56$--$0.72$, the dynamical diagnostics show a smooth transition from a decelerated to an accelerated phase. While the $Om(z)$ diagnostic exhibits a negative slope, indicating that the model is not $\Lambda$CDM, the effective equation-of-state parameter $\omega_{\text{eff}}(z)$ stays within the quintessence regime ($-1 < \omega_{\text{eff}} < -1/3$). The analysis of classical energy conditions shows that the WEC, DEC, and NEC are satisfied throughout the cosmic evolution, with a violation of the SEC at lower-redshift, which is consistent with late-time acceleration. Linear homogeneous perturbation analysis further confirms the model's dynamical stability. Conclusively, the FHDE-inspired reconstructed $f(Q)$ gravity provides a stable, observationally compatible, and geometrically motivated alternative to $\Lambda$CDM, that successfully describes the late-time cosmic acceleration within the symmetric teleparallel framework.
\end{abstract}

	
	\maketitle

\section{Introduction}\label{sec1}
One of the most significant turning points in contemporary world of cosmology is the realisation that the universe is experiencing a phase of late-time accelerated expansion. Initially discovered through observations of Type Ia Supernovae (SNeIa) \cite{Riess1998, Perlmutter1999, Astier2006}, this remarkable feature has since been confirmed by independent probes, such as high-precision measurements of the Cosmic Microwave Background Radiation (CMBR) \cite{Spergel2003}, Baryon Acoustic Oscillations (BAO), and other large-scale structure (LSS) surveys \cite{Tegmark2004, Cole2005, Eisenstein2005}. Collectively, these observations provide a vital compelling evidence that the universe is not only expanding but also under going this expansion at an accelerated pace. Although the General Relativity (GR), in conjunction with the homogeneous and isotropic Friedmann–Lemaître–Robertson–Walker (FLRW) framework, have been incredibly successful in explaining cosmic evolution on large scales, the theoretical source of this accelerated expansion is still unknown. Two separate epochs of acceleration are needed in the standard paradigm: firstly an early-time inflationary era, often modelled through scalar field dynamics, and secondly the present-day acceleration, which is most commonly attributed to a cosmological constant within the Einstein field equations. Together, these ingredients define the $\Lambda$CDM model, which has been successfully consistent with observational datasets across different epochs of cosmic history. However, despite its empirical success, the $\Lambda$CDM framework suffers from deep theoretical challenges. The most prominent among them is the cosmological constant problem, arising from the enormous discrepancy between the theoretically predicted and the observationally inferred values of vacuum energy density \cite{Weinberg1989, Martin2012}. Closely related are the fine-tuning and coincidence problems, which question why the observed value of dark energy density is so finely adjusted and why its dominance coincides with the current cosmological epoch. These conceptual difficulties have motivated extensive research into alternative approaches either by invoking dynamical dark energy models with evolving equations of state or by considering modifications of Einstein’s theory of gravity itself. A comprehensive overview of these developments can be found in Refs.~\cite{za1}--\cite{za10}.\\
In recent years, a wide variety of dynamical dark energy models have been proposed in an attempt to overcome the conceptual difficulties associated with the $\Lambda$CDM paradigm. Among these, the holographic dark energy (HDE), based on the holographic principle of quantum gravity, has attracted significant attention. The basic idea of HDE is that the vacuum energy density of the universe can be constrained through an infrared (IR) cutoff of the system, typically linked to a cosmological horizon. A number of generalisations have emerged from this framework, such as agegraphic dark energy (ADE) \cite{Cai2007}, new agegraphic dark energy (NADE) \cite{Wei2007}, and more recently, \textit{fractional holographic dark energy} (FHDE) \cite{Trivedi2024}. A more general description of vacuum energy density with possible connections to non-local and quantum gravitational effects is provided by the FHDE model, which integrates fractional order calculus into the HDE framework \cite{Trivedi2024, Rathore2025, Bidlan2025}. Depending on the choice of IR cutoff, commonly taken as the Hubble horizon in recent studies, FHDE leads to different cosmic dynamics, which can be tested against observations. Its flexibility and broader theoretical motivation make FHDE an compelling candidate for exploring late-time cosmic acceleration beyond the conventional constant dark energy paradigm \cite{Rathore2025, Bidlan2025}. Simultaneously, an alternative approach to explain the observed late-time acceleration involves the modified theories of gravity, which generalise the gravitational sector by adding geometric contributions instead of new additional dark energy components. Such theories can reproduce consistent cosmological dynamics and remain compatible with current observational data, often without the need for an explicit dark energy term \cite{Capozziello2006, Nojiri2006recon, Sotiriou2010}. Among these frameworks, the recently developed \textit{$f(Q)$ gravity} models have attracted considerable attention. Unlike curvature-based $f(R)$ or torsion-based $f(T)$ gravities, $f(Q)$ gravity is formulated within the symmetric teleparallel geometry, where gravitational effects are entirely encoded in the non-metricity scalar $Q$ \cite{Jimenez2018, Mandal2020}. This formulation leads to second-order field equations, thereby simplifying the dynamics, and also avoids the local Lorentz violation and antisymmetric contributions that are typically encountered in $f(T)$ theories.\\
Literature shows one of the most effective strategies to tackle the cosmological constant problem and to model late-time cosmic acceleration within modified gravity frameworks is the \textit{cosmological reconstruction method}. This approach aims to determine a suitable functional form of the gravitational action that can reproduce a given expansion history or a phenomenologically favored cosmological evolution, such as that observed in the $\Lambda$CDM model. By relying on the observed or desired dynamics rather than postulating an explicit form of the gravity function, reconstruction provides a bottom-up method to infer the underlying gravitational theory from cosmic evolution. Nevertheless, due to the inherent non-linearity and complexity of the field equations in most modified gravity theories, obtaining exact analytical solutions is often challenging, and numerical solutions can be computationally demanding and sensitive to model parameters. The reconstruction methodology typically involves inverting the cosmological dynamics in a flat FLRW background to identify classes of modified gravity theories, such as $f(R)$, $f(G)$, $f(T)$, $f(Q)$, or $f(R,T)$ that can mimic the desired cosmological behavior \cite{Nojiri2006recon, Nojiri2011rev, Dunsby2010, Goheer2009, Wu2010, Bamba2011recon, Baffou2017, Harko2018, Mandal2020}. Historically, the first systematic applications of this approach were developed for $f(R)$ gravity, where Nojiri and Odintsov \cite{Nojiri2006recon, Nojiri2011rev} effectively showed how observationally motivated expansion histories could be used to reconstruct viable $f(R)$ functions, establishing a strong connection between cosmic evolution and the gravitational Lagrangian. Later studies by Dunsby et al. \cite{Dunsby2010} showed how difficult it is to precisely replicate the $\Lambda$CDM scenario within $f(R)$ frameworks, suggesting that effective matter contributions may be required in certain situations. Later, other geometric frameworks were added to the reconstruction paradigm. In $f(G)$ gravity, Goheer et al. \cite{Goheer2009} demonstrated that only specific classes of models in $f(G)$ gravity can admit exact power-law solutions, highlighting the possibilities and limitations of reconstruction in Gauss–Bonnet extensions. Around the same time teleparallel theories got comparable attention. Wu \& Yu \cite{Wu2010} and Bamba et al. \cite{Bamba2011recon} demonstrated in $f(T)$ gravity how parametrisations of the Hubble parameter or scale factor in appropriate ways can result in models that can unify several cosmological epochs, including the late-time accelerated phase. Baffou et al. \cite{Baffou2017} investigated further generalisations that included explicit couplings between torsion and matter fields, such as $f(\tau,T)$ gravity. With the introduction of symmetric teleparallel geometry, cosmological reconstruction entered a new stage or phase in $f(Q)$ gravity. The consistency, energy conditions, and dynamical viability of $f(Q)$ cosmologies were deeply investigated in foundational works by Harko et al. \cite{Harko2018} and Mandal et al. \cite{Mandal2020}. More recently, using infrared cutoffs motivated by holographic dark energy models, authors such as Saha and Rudra \cite{Saha2025} proposed holographic reconstructions of $f(Q)$ gravity. When taken as a whole, these advancements show how reconstruction techniques have clearly evolved, from curvature-based $f(R)$ theories, extending to $f(G)$ and torsion-based $f(T)$ frameworks, incorporating matter couplings, and finally adapting to symmetric teleparallel $f(Q)$ gravity. This highlights the ability of reconstruction methods to build phenomenologically viable modified gravity models that provide alternative pathways to the theoretical difficulties present in the conventional $\Lambda$CDM paradigm while also being consistent with observational data.\\
In this work, utilizing a direct correspondence with Fractional Holographic Dark Energy (FHDE), we focus on exploring a cosmological reconstruction scheme within the framework of symmetric teleparallel gravity or $f(Q)$ gravity. As literature suggests integration of fractional calculus into the holographic principle, giving rise to FHDE. Which is a generalisation of conventional holographic dark energy models, naturally integrates quantum gravitational effects into the cosmological dynamics \cite{Trivedi2024, Rathore2025, Bidlan2025}. Because of this FHDE becomes particularly a  appealing candidate for reconstructing modified gravity models that seek to explain late-time acceleration. From literature previous investigations have shown that reconstruction methods employing various dark energy models, including standard HDE, new agegraphic dark energy (NADE), and their entropy-corrected extensions, can successfully give rise to viable forms of modified gravity functions \cite{Saha2025, Nojiri2006recon}. When implemented within frameworks such as $f(R)$, $f(T)$, or $f(Q)$ gravity, have yielded models that reproduce the key aspects of cosmic evolution while also being consistent with observational data. Inspired by these advances, we extend this methodology to $f(Q)$ gravity, which offers a dynamically flexible and geometrically rich substitute for theories based on curvature and torsion. Primarily our main goal is to use the FHDE energy density as input to reconstruct the functional form of $f(Q)$ and analyse the resulting cosmological behaviour that it predicts. The theoretical significance of this work is further enhanced by the fractional and quantum-inspired origin of FHDE, which may provide fresh perspectives on how quantum cosmology and modified gravity interact. It is worth emphasizing how our work differs from previous reconstruction studies, like Ref.\cite{Saha2025}, which used holographic dark energy models like Granda–Oliveros and Chen–Jing with particular infrared cutoffs. On the other hand, our reconstruction offers a more flexible and general framework by using FHDE with the Hubble horizon as infrared cutoffs. This method ensures better compatibility with observational constraints, richer dynamical evolution, and causal consistency. In summary, our study presents a complementary yet distinct reconstruction scheme that unifies the strengths of FHDE with the symmetric teleparallel formulation of $f(Q)$ gravity, offering a promising avenue for addressing late-time cosmic acceleration within a quantum-motivated framework.\\
This paper is organised as follows: A brief mathematical formalism of $f(Q)$ gravity and the cosmological framework in a flat FLRW Universe is given in Sec. \ref{s2}, and the reconstruction for $f(Q)$ gravity using FHDE is described in Sec. \ref{s3}. The methodology and observational data utilised in the study to constrain the model are covered in Sec. \ref{s4}. The models' constraints and the cosmological implications they reflect are covered in Sec \ref{s5}. Lastly, a summary of the study is given in Sec. \ref{c1}.

\section{$f(Q) Gravity$}\label{s2}
The theoretical framework of $f(Q)$ gravity, in which the gravitational interaction is attributed to the non-metricity scalar $Q$ rather than the curvature scalar $R$ or the torsion scalar $T$, is briefly presented in this section. In other words, in $f(Q)$ gravity, it is assumed that the connection has non-vanishing non-metricity but is free of curvature and torsion \cite{Jimenez2018}.
\begin{equation}
Q^\gamma_{\mu\nu} \equiv -\nabla_\gamma g_{\mu\nu}.
\end{equation}
This non-metricity measures the change in a vector's length under parallel transport. The affine connection can be broken down as follows:
\begin{equation}
\Upsilon^\gamma_{\mu\nu} = \Gamma^\gamma_{\mu\nu} + L^\gamma_{\mu\nu},
\end{equation}
where $\Gamma^\gamma_{\mu\nu}$ is the Levi-Civita connection and $L^\gamma_{\mu\nu}$ is the disformation tensor constructed from the non-metricity tensor. The non-metricity scalar $Q$, defined as follows, is obtained by contracting the non-metricity tensor with a superpotential tensor $P^\gamma_{\mu\nu}$.
\begin{equation}
Q = -Q^\gamma_{\mu\nu} P^\mu_\gamma{}^\nu.
\end{equation}
The action for $f(Q)$ gravity in the presence of matter fields is given by \cite{Jimenez2018}:
\begin{equation}
S = \int d^4x \sqrt{-g} \left[ -\frac{1}{2} f(Q) + \mathcal{L}_m \right],
\end{equation}
where $\mathcal{L}_m$ represents the matter Lagrangian and $f(Q)$ is a general function of the non-metricity scalar. The modified gravitational field equations are obtained by varying the action with respect to the metric:
\begin{equation}
2 \nabla_\gamma \left( \sqrt{-g} f_Q P^\gamma_{\mu\nu} \right) + \sqrt{-g} \left[ \frac{1}{2} f g_{\mu\nu} + f_Q \left(P_{\nu\rho\sigma} Q_\mu{}^{\rho\sigma} - 2 P_{\rho\sigma\mu} Q^\rho{}_{\nu}{}^\sigma \right) \right] = \sqrt{-g} T_{\mu\nu},
\end{equation}
where $T_{\mu\nu}$ is the energy-momentum tensor and $f_Q = df/dQ$. Spacetime becomes flat and calculations are made easier in the coincident gauge, where the connection disappears ($\Upsilon^\gamma_{\mu\nu} = 0$). The covariant derivatives in this gauge reduce to partial derivatives since the non-metricity tensor becomes $Q^\gamma_{\mu\nu} = -\partial_\gamma g_{\mu\nu}$.\\
A spatially flat Friedmann Lemaître Robertson Walker (FLRW) metric is assumed in the study, in order to explore cosmological implications, given as:
\begin{equation}
ds^2 = -dt^2 + a^2(t)\left( dx^2 + dy^2 + dz^2 \right),
\end{equation}
where $a(t)$ represents the scale factor. For this metric, the non-metricity scalar becomes:
\begin{equation}
Q = 6H^2,
\end{equation}
where $H = \dot{a}/a$ is the Hubble parameter. We study the energy-momentum tensor, $T_{\mu\nu} = (\rho + p)u_\mu u_\nu + p g_{\mu\nu}$, where $u^\mu = (-1,0,0,0)$ is the four-velocity of the fluid that satisfies the requirement $u^\mu u_\mu = -1$. The pressure and energy density are denoted here by $\rho$ and $p$, respectively. The following are the modified Friedmann equations in $f(Q)$ gravity with a perfect fluid energy-momentum tensor and the previously mentioned metric:
\begin{equation}
3H^2 = \frac{1}{2f_Q} \left( \rho + \frac{f}{2} \right), \qquad \dot{H} + 3H^2 + \frac{\dot{f}_Q}{f_Q} H = \frac{1}{2f_Q} \left( -p + \frac{f}{2} \right).
\label{x8}
\end{equation}
In the above equations the derivative with respect to time is represented by the overhead dot (.) and we make the assumption that $8\pi G = c = 1$. The aforementioned formulas explain how the cosmological constant, spatial curvature, and the energy density of matter and radiation affect the universe's expansion. Putting $f(Q) = Q$ restores the standard GR equations. But when $f(Q)=Q+F(Q)$ is inserted, the field equations become:
\begin{equation}
3H^2 = \rho + F - 2QF_Q,
\end{equation}
\begin{equation}
\left(2QF_{QQ} + F_Q + 1\right) \dot{H} + \frac{1}{4}\left(Q + 2QF_Q - F\right) = -2p,
\end{equation}
where \( F_Q = \frac{dF}{dQ} \) and \( F_{QQ} = \frac{d^2F}{dQ^2} \). These equations can be further defined as follows in the context of isotropic and homogeneous spatially flat FLRW cosmology:
\begin{equation}
3H^2 = \rho_m + \rho_r + \rho_{de}, \qquad 2\dot{H} + 3H^2 = -(p_m + p_r + p_{de}).
\end{equation}
Here, the energy densities of the radiation and matter components are denoted by $\rho_r$ and $\rho_m$, respectively, while the pressure corresponding to the matter and radiation components is denoted by $p_m$ and $p_r$. The pressure $p_{de}$ and effective dark energy density $\rho_{de}$ resulting from the geometric contribution may be expressed as \cite{Nojiri2006recon,Karami2011a}:
\begin{equation}
\rho_{de} = \frac{F}{2} - Q F_Q,
\label{e12}
\end{equation}
\begin{equation}
p_{de} = 2 \dot{H} \left(2Q F_{QQ} + F_Q\right) - \rho_{de},
\label{x13}
\end{equation}
Such geometric terms with an effective dark-energy sector are identified in accordance with standard methods in modified gravity reconstructions \cite{Nojiri2006recon,Karami2011a}. Additionally, as demonstrated above \cite{Saha2025,Nojiri2006recon}, the reconstructed $f(Q)$ model satisfies the fundamental viability requirements of $f(Q)$ gravity along the best-fit background, namely $1+F_{Q}>0$ (positive effective coupling) and $F_{QQ}\geq 0$ (absence of ghost-like instabilities), in accordance with Eqs.~\ref{x8})–(\ref{x13}). Their energy conservation equations, taking into account non-interacting radiation, matter, and geometric dark energy components, are:
\begin{equation}
\dot{\rho}_r + 4H\rho_r = 0, \qquad \dot{\rho}_m + 3H\rho_m = 0, \qquad \dot{\rho}_{de} + 3H(1 + \omega_{de})\rho_{de} = 0.
\end{equation}
where, $\omega_{de}=\frac{p_{de}}{\rho_{de}}$. These can help to derive standard redshift evolution:
\begin{equation}
\rho_r(z) = \rho_{r0}(1 + z)^4, \qquad \rho_m(z) = \rho_{m0}(1 + z)^3.
\end{equation}
where $\rho_{r0}$ and $\rho_{m0}$ denote the present values of $\rho_{r}$ and $\rho_{r}$.

\section{Reconstruction of \boldmath{$f(Q)$} Gravity from FHDE}
\label{s3}
In this section, we provide the reconstruction methodology that uses the Fractional Holographic Dark Energy (FHDE) model as input to determine the functional form of $f(Q)$ gravity. The FHDE model, recently proposed in the context of fractional cosmology \cite{Trivedi2024, Rathore2025, Bidlan2025}, generalizes the standard holographic dark energy (HDE) scenario by employing a fractional modification of the holographic principle. The energy density of FHDE can be written as:
\begin{equation}
\rho_{\text{FHDE}} = 3 c^2 M_P^2 L^{\frac{2-3\alpha}{\alpha}},
\label{e16}
\end{equation}
where $c$ is a dimensionless model parameter, $M_P^2 = 1/\kappa^2$ is the reduced Planck mass squared, $L$ denotes the chosen infrared (IR) cutoff length, and $\alpha$ is the fractional parameter with the constraint $1<\alpha \leq 2$. For $\alpha=2$, Eq.~\eqref{e16} reduces to the standard HDE expression. Before proceeding with the reconstruction, it is necessary to fix the choice of the infrared (IR) cutoff $L$ in the FHDE density. In the literature, several cutoffs have been proposed, each with distinct physical motivations. The earliest suggestion was the Hubble horizon cutoff, $L = H^{-1}$, which introduces a natural length scale associated with the cosmic expansion rate and was originally intended to address the fine-tuning problem. However, in the standard HDE framework this choice leads to a nearly vanishing equation of state (EoS) parameter for dark energy, thereby failing to drive the observed late-time acceleration \cite{Granda2008}. Alternative possibilities include the particle horizon, the future event horizon, the Granda–Oliveros cutoff incorporating $\dot{H}$, and the generalized Nojiri–Odintsov cutoff which can depend on combinations of $H$, $\dot{H}$, and higher derivatives \cite{Granda2008, Kim2013, Nojiri2017, Nojiri2019}. Among these, the future event horizon has been widely employed because it naturally yields accelerated expansion, though it suffers from causality issues, while generalized cutoffs offer flexibility but lack universal physical motivation \cite{Kim2013, Li2004}. Moreover, all of these cutoffs face challenges related to classical stability under perturbations \cite{Li2004, Myung2007}. However, in the context of FHDE, the situation is significantly improved. Recent studies \cite{Trivedi2024, Rathore2025, Bidlan2025} have shown that the inclusion of fractional entropy corrections modifies the scaling of the energy density in such a way that even with the Hubble horizon cutoff, the FHDE model can yield viable late-time cosmic acceleration. In particular, smaller values of the fractional parameter $\alpha$ provide acceptable cosmological evolution, including realistic behaviour for the density parameters, the deceleration parameter, and the EoS of dark energy \cite{Trivedi2024}. This is a notable improvement over conventional HDE, where the Hubble horizon cutoff is usually discarded. The advantage of adopting $L=H^{-1}$ in FHDE is therefore three-fold: (i) it ensures causally consistent dynamics without reference to the future evolution of the scale factor, (ii) it offers a simple and direct link between the fractional holographic density and the non-metricity scalar $Q=6H^2$, which is essential for reconstructing $f(Q)$ gravity in closed form, and (iii) recent studies of FHDE with the Hubble horizon as the cutoff have been seen to provide an acceptable evolution. For these reasons, we shall adopt the Hubble horizon cutoff in the present reconstruction analysis, while noting that extensions to more generalized cutoffs may be explored in future work.\\
Combining Eq. (\ref{e16}) and Eq. (\ref{e12}) as shown below, for $F(Q)$ we obtain a differential equation:
\begin{equation}
 F(Q) - 2Q F_Q = 6 c^{2} L^{\frac{2-3\alpha}{\alpha}}.
\label{eq:recon_diff}
\end{equation}
where $\kappa^2 = 1$ has been used. This linear differential equation in $F(Q)$ is of first order. Using the Hubble horizon cutoff ($L=H^{-1}$) in Eq. (\ref{eq:recon_diff}), we now obtain:
\begin{equation}
 F(Q) - 2Q F_Q = 6 c^{2} H^{\frac{3\alpha-2}{\alpha}}.
\label{eq:recon_diff_2}
\end{equation}
Now, we know that in symmetric teleparallel gravity
\begin{equation}
Q = 6H^2 \quad\Longrightarrow\quad H=\sqrt{\frac{Q}{6}}.
\end{equation}
Substituting this relation into Eq.~\eqref{eq:recon_diff_2} gives us:
\begin{equation}
F(Q) - 2Q F_Q \;=\; 6 c^2 \left(\sqrt{\frac{Q}{6}}\right)^{\frac{3\alpha-2}{\alpha}}.
\end{equation}
Solving the above differential equation, we obtain the reconstructed $f(Q)$ gravity as:
\begin{equation}
f(Q)=Q+F(Q), \quad
F(Q) = \frac{6^{1-\frac{3 \alpha -2}{2 \alpha }} c^2 Q^{\frac{3 \alpha -2}{2 \alpha }}}{1-\frac{3 \alpha -2}{\alpha }}+C_1 \sqrt{Q},
\label{e23}
\end{equation}
where, $C_1$ is some arbitrary constant of integration. Here, for simplicity of calculation and representation we can introduce another constant say $\gamma=\frac{3 \alpha -2}{2 \alpha }$, using it we have the reconstructed $f(Q)$ gravity as:
\begin{equation}
f(Q)=Q+F(Q), \quad
F(Q) = \frac{6^{1-\gamma } c^2 Q^{\gamma }}{1-2 \gamma }+C_1 \sqrt{Q},
\label{e23}
\end{equation}
Here, initially as the fractional parameter $\alpha$ was constrained with $1<\alpha \leq 2$, so $\gamma=\frac{3 \alpha -2}{2 \alpha }$ will have the constrain $\frac{1}{2}<\gamma \leq 1$. Now, using the condition $f_{Q}(Q_{0}) = 1$, where $Q_{0} = 6H_{0}^{2}$ is the present-day value of the non-metricity scalar, we can determine the integration constant $C_{1}$. This yields
\begin{equation}
C_{1} = \frac{2^{2-\gamma } 3^{1-\gamma } c^2 \gamma  Q_0^{\gamma -\frac{1}{2}}}{2 \gamma -1}
\label{eq:C1_condition}
\end{equation}
So, finally using this in Eq. (\ref{e23}) we obtain the reconstructed $f(Q)$ gravity as:
\begin{equation}
f(Q)=Q+F(Q), \quad
F(Q) = \frac{6^{1-\gamma } c^2 Q^{\gamma } \left(1-2 \gamma  \left(\frac{Q}{Q_0}\right)^{\frac{1}{2}-\gamma }\right)}{1-2 \gamma },
\label{e23a}
\end{equation}
where the integration constant has been fixed by imposing the present-day normalization condition $f_{Q}(Q_{0})=1$, with $Q_0 = 6H_0^2$. This choice ensures that the effective gravitational coupling reduces to the Newtonian value at the current epoch, thereby satisfying local gravity constraints. The parameter $\gamma$ is restricted as $\frac{1}{2}<\gamma\le 1$, corresponding to the original constraint on the fractional parameter as $1<\alpha\le 2$, which guarantees the consistency of the reconstructed function and avoids singularities in the denominator. The additional term $F(Q)$ encapsulates deviations from standard General Relativity, allowing for late-time acceleration or other cosmological effects depending on the choice of parameters $c$ and $\gamma$. Importantly, in the limit $c \to 0$, the correction term vanishes, and the theory smoothly reduces to $f(Q)=Q$, thus recovering General Relativity as expected. This shows that the reconstructed $f(Q)$ model can both accommodate modifications to cosmology and remain consistent with standard gravitational physics in the appropriate limit. In the following section we will discuss and evaluate the constraints on the model parameters using observational data.
\section{Observational Data and Methodology for Constraints}\label{s4}
Complementary compilations of the Hubble parameter $H(z)$, such as the recent DESI DR2 BAO dataset \cite{ref98}, a previous compilation of BAO measurements (P-BAO) including SDSS, WiggleZ, and other surveys \cite{d113}-\cite{d79}, and Cosmic Chronometers, which are measurements of H(z) from the relative ages of passively-evolving galaxies \cite{x38}, as described in Table \ref{tab1} are used to obtain the constraint on the parameters of the reconstructed $f(Q)$ gravity model. There are no overlapping bins among these samples, and their redshift coverage is complimentary. We regard all $H(z)$ values as uncorrelated as the constituent surveys are independent. Therefore, without using a covariance matrix, the total $\chi^{2}$ is calculated as the sum of the individual $\chi^{2}$ contributions from all data points. Although other cosmic probes, including Type Ia supernovae, can also be utilised for this purpose, we limit our attention to the aforementioned three datasets for job simplicity. We use a Markov Chain Monte Carlo (MCMC) framework to analyse the data and derive strong restrictions on the model and cosmological parameters. In light of this, important cosmological indicators including the deceleration parameter ($q(z)$), the effective equation of state ($\omega_{eff}(z)$), $om(z)$ diagnostics, and ultimately the energy conditions and stability analysis are used to study the cosmological scenarios.\\
Literature shows that the modified Friedmann equation can be recast for this model as:
\begin{equation}
\left(\frac{H(z)}{H_0}\right)^{2} = \Omega_{r0} (1 + z)^4 + \Omega_{m0} (1 + z)^3 + \frac{1}{3H_0^2} \left[ \frac{1}{2}F(Q(z)) - Q(z) F_Q(Q(z)) \right],
\label{e29}
\end{equation}
In this case, the Hubble parameter's present value is $H_0$, whereas the radiation and matter density parameter's present values are denoted by $\Omega_{r0}$ and $\Omega_{m0}$, respectively. This approach allows us to directly constrain the reconstructed $f(Q)$ model through its FHDE-driven contribution while comparing it with observational Hubble data via standard statistical techniques. Additionally, after obtaining $H(z)$, we may get the deceleration parameter, which is as follows:
\begin{equation}
    q(z) = -1 - \frac{\dot{H}}{H^2} = -1 + (1+z) \frac{1}{H(z)} \frac{dH}{dz},
    \label{eq}
\end{equation}
It shows how rapidly the universe is expanding or contracting. Accelerated expansion is indicated by a negative value of $q(z)$. Additionally, we may get the effective EoS, which is as follows:
\begin{equation}
\omega_{eff}(z) = \frac{p_{eff}}{\rho_{eff}}=-1 + \frac{2(1+z)}{3H(z)}\frac{dH}{dz},
 \label{ew}
\end{equation}
This, with $p_{eff}$ and $\rho_{eff}$ representing the effective pressure and energy density, offers an efficient fluid account of the dynamics of the Universe that incorporates both matter and dark energy effects. These values enable us to evaluate if the effective equation of state crosses the phantom boundary ($w_{\text{eff}} < -1$) and whether the model predicts a change from deceleration to acceleration. In addition, we have the $Om(z)$ diagnostic, which has the following definition:
\begin{equation}
Om(z) = \frac{H^2(z)/H_0^2 - 1}{(1 + z)^3 - 1}.
\label{eo}
\end{equation}
It is a purely geometrical tool made from the Hubble parameter that makes it possible to distinguish between dark energy theories without needing the knowledge of the equation of state.\\
For our model using Eq. (\ref{eq:recon_diff_2}) in Eq. (\ref{e29}) we obtain:
\begin{equation}
\left(\frac{H(z)}{H_0}\right)^{2} = \Omega_{r0} (1 + z)^4 + \Omega_{m0} (1 + z)^3 + \frac{1}{3H_0^2} \left[ c^{2} H(z)^{\frac{3\alpha-2}{\alpha}}\right],
\label{e29a}
\end{equation}
As done previously defining $\gamma=\frac{3 \alpha -2}{2 \alpha }$, helps us to re-write the above equation as:
\begin{equation}
\left(\frac{H(z)}{H_0}\right)^{2} = \Omega_{r0} (1 + z)^4 + \Omega_{m0} (1 + z)^3 + \frac{1}{3H_0^2} \left[ c^{2} H(z)^{2\gamma}\right],
\label{e29a}
\end{equation}
In the above Eq. (\ref{e29a}) using the condition $H(z)=H_0$ at $z=0$, we can obtain the parameter $c^{2}$ as:
\begin{equation}
c^{2}=3 H_0^{2-2 \gamma } \left(1-\Omega _{m0}-\Omega _{r0}\right)
\label{e29b}
\end{equation}
So, using Eqs. (\ref{e29a}) and (\ref{e29b}) we obtain:
\begin{equation}
\left(\frac{H(z)}{H_0}\right)^{2} = \Omega_{r0} (1 + z)^4 + \Omega_{m0} (1 + z)^3 + \left(\frac{H}{H_0}\right)^{2\gamma}\left(1-\Omega _{m0}-\Omega _{r0}\right),
\label{e29c}
\end{equation}
Equation~(\ref{e29c}) is an implicit equation for $H(z)$. Hence, for given model parameters $\theta = \{\gamma, \Omega_{m0}, H_0\}$, we solve this equation numerically at each redshift $z_i$ to obtain the theoretical Hubble values $H_{th}(z_i)$. These values are then compared with observational Hubble datasets to find the best fitting parameter values for the given model using the standard chi-squared minimization method. For which the chi-squared function is defined as:
\begin{equation}
\chi^2_H(\theta) = \sum_{i=1}^{N} \frac{\left[H_{th}(z_i,\theta) - H_{obs}(z_i)\right]^2}{\sigma_H(z_i)^2},
\label{eq:chi2_H}
\end{equation}
where $H_{obs}(z_i)$ and $\sigma_H(z_i)$ denote the measured Hubble parameter and its uncertainty at redshift $z_i$. A Markov Chain Monte Carlo (MCMC) method is then applied to explore the parameter space and determine the posterior distributions of $\gamma$, $\Omega_{m0}$, and $H_0$. We use the following priors on the model parameters for the MCMC analysis: $H_0 \in [60, 80]$, $\Omega_{m0} \in [0, 1]$, and $\gamma \in [0.5, 1]$. The parameter values $(H_0, \Omega_{m0}, \gamma) = (68, 0.3, 0.6)$ were used to initialise the MCMC chains. For simplicity, the radiation density is set at $\Omega_{r0}=0.0003$ throughout the study.\\
We conduct the MCMC analysis in three stages: using the DESI dataset alone, then with the previous BAO dataset and DESI dataset, and finally with a combination previous BAO dataset, DESI dataset and Cosmic Chronometers. Statistical indicators like the coefficient of determination ($R^2$), the minimum chi-squared value ($\chi^2_{\min}$), the Akaike Information Criterion (AIC), and the Bayesian Information Criterion (BIC) are used to compare the results with those from the standard $\Lambda$CDM model (see Refs.~\cite{refS66}, \cite{refS68}, \cite{refS64} for details).Because of the displays of a better balance between complexity and quality of fit, a model with lower AIC, BIC, and $\chi^2_{\min}$ values is statistically preferred. These include $R^2$ and $\chi^2_{\min}$, which evaluate fit quality without taking the number of free parameters into consideration. On the other hand, models with more complexity are penalised by both AIC and BIC; however, BIC imposes a more severe penalty, particularly when dealing with bigger datasets. We calculate the differences in AIC and BIC as follows to assess the relative performance of models: \( \Delta X = \Delta \text{AIC} \) or \( \Delta X = \Delta \text{BIC} \). The following is how these differences are comprehended:
\begin{itemize} 
\item \textbf{$0 \leq \Delta X \leq 2$}:It is impossible to determine whether the model is better because the evidence is \textit{weak}.

\item \textbf{$2 < \Delta X \leq 6$}: The model with the lower value is supported by \textit{positive} evidence. 
\item \textbf{$6 < \Delta X \leq 10$}: The evidence is regarded as \textit{strong}.
\end{itemize}
In order to ensure accuracy and precession in cosmological inference, these statistical diagnostics are crucial requirements for model comparison.
\begin{table}[h!]
\centering
\begin{tabular}{|ccc|c||ccc|c||ccc|c|}
\hline
\multicolumn{4}{|c||}{\textbf{DESI}} & \multicolumn{4}{c||}{\textbf{P-BAO}} & \multicolumn{4}{c|}{\textbf{CC}} \\
\hline
$z$ & $H(z)$ & $\sigma_H$ & Ref & $z$ & $H(z)$ & $\sigma_H$ & Ref & $z$ & $H(z)$ & $\sigma_H$ & Ref \\
\hline
0.51 & 97.21 & 2.83 & \cite{ref98} & 0.24 & 79.69 & 2.99 & \cite{d113} & 0.07 & 69.00 & 19.60 & \cite{cc54} \\
0.71 & 101.57 & 3.04 & \cite{ref98} & 0.30 & 81.70 & 6.22 & \cite{d114} & 0.09 & 69.00 & 12.00 & \cite{cc55} \\
0.93 & 114.07 & 2.24 & \cite{ref98} & 0.31 & 78.17 & 6.74 & \cite{d115} & 0.12 & 68.60 & 26.20 & \cite{cc54} \\
1.32 & 147.58 & 4.49 & \cite{ref98} & 0.34 & 83.17 & 6.74 & \cite{d113} & 0.17 & 83.00 & 8.00 & \cite{cc55} \\
2.33 & 239.38 & 4.80 & \cite{ref98} & 0.35 & 82.70 & 8.40 & \cite{d116} & 0.179 & 75.00 & 4.00 & \cite{cc56} \\
 &  &  &  & 0.36 & 79.93 & 3.39 & \cite{d115} & 0.199 & 75.00 & 5.00 & \cite{cc56} \\
 &  &  &  & 0.38 & 81.50 & 1.90 & \cite{d5} & 0.20 & 72.90 & 29.60 & \cite{cc54} \\
 &  &  &  & 0.40 & 82.04 & 2.03 & \cite{d115} & 0.27 & 77.00 & 14.00 & \cite{cc55} \\
 &  &  &  & 0.43 & 86.45 & 3.68 & \cite{d113} & 0.28 & 88.80 & 36.60 & \cite{cc54} \\
 &  &  &  & 0.44 & 82.60 & 7.80 & \cite{d74} & 0.352 & 83.00 & 14.00 & \cite{cc56} \\
 &  &  &  & 0.44 & 84.81 & 1.83 & \cite{d115} & 0.3802 & 83.00 & 13.50 & \cite{cc58} \\
 &  &  &  & 0.48 & 87.79 & 2.03 & \cite{d115} & 0.4 & 95.00 & 17.00 & \cite{cc55} \\
 &  &  &  & 0.56 & 93.33 & 2.32 & \cite{d115} & 0.4004 & 77.00 & 10.20 & \cite{cc58} \\
 &  &  &  & 0.57 & 87.60 & 7.80 & \cite{d10} & 0.4247 & 87.10 & 11.20 & \cite{cc58} \\
 &  &  &  & 0.57 & 96.80 & 3.40 & \cite{d117} & 0.4497 & 92.80 & 12.90 & \cite{cc58} \\
 &  &  &  & 0.59 & 98.48 & 3.19 & \cite{d115} & 0.47 & 89.00 & 50.00 & \cite{cc59} \\
 &  &  &  & 0.60 & 87.90 & 6.10 & \cite{d74} & 0.4783 & 80.90 & 9.00 & \cite{cc58} \\
 &  &  &  & 0.61 & 97.30 & 2.10 & \cite{d5} & 0.48 & 97.00 & 62.00 & \cite{cc59} \\
 &  &  &  & 0.64 & 98.82 & 2.99 & \cite{d115} & 0.593 & 104.00 & 13.00 & \cite{cc56} \\
 &  &  &  & 0.978 & 113.72 & 14.63 & \cite{d118} & 0.68 & 92.00 & 8.00 & \cite{cc56} \\
 &  &  &  & 1.23 & 131.44 & 12.42 & \cite{d118} & 0.781 & 105.00 & 12.00 & \cite{cc56} \\
 &  &  &  & 1.48 & 153.81 & 6.39 & \cite{d79} & 0.875 & 125.00 & 17.00 & \cite{cc56} \\
 &  &  &  & 1.526 & 148.11 & 12.71 & \cite{d118} & 0.88 & 90.00 & 40.00 & \cite{cc59} \\
 &  &  &  & 1.944 & 172.63 & 14.79 & \cite{d118} & 0.9 & 117.00 & 23.00 & \cite{cc55} \\
 &  &  &  & 2.30 & 224.00 & 8.00 & \cite{d119} & 1.037 & 154.00 & 20.00 & \cite{cc56} \\
 &  &  &  & 2.36 & 226.00 & 8.00 & \cite{d120} & 1.3 & 168.00 & 17.00 & \cite{cc55} \\
 &  &  &  & 2.40 & 227.80 & 5.61 & \cite{d121} & 1.363 & 160.00 & 33.60 & \cite{cc60} \\
 &  &  &  &  &  &  &  & 1.43 & 177.00 & 18.00 & \cite{cc55} \\
 &  &  &  &  &  &  &  & 1.53 & 140.00 & 14.00 & \cite{cc55} \\
 &  &  &  &  &  &  &  & 1.75 & 202.00 & 40.00 & \cite{cc55} \\
 &  &  &  &  &  &  &  & 1.965 & 186.50 & 50.00 & \cite{cc60} \\
\hline
\end{tabular}
\caption{Hubble parameter $H(z)$ and its uncertainty at redshift $z$ from the Cosmic Chronometer (CC), DESI, and P-BAO datasets, measured in units of $km s^{-1} Mpc^{-1}$.}
\label{tab1}
\end{table}
\section{Constraints and Cosmological Implications}\label{s5}
\begin{figure}[htb]
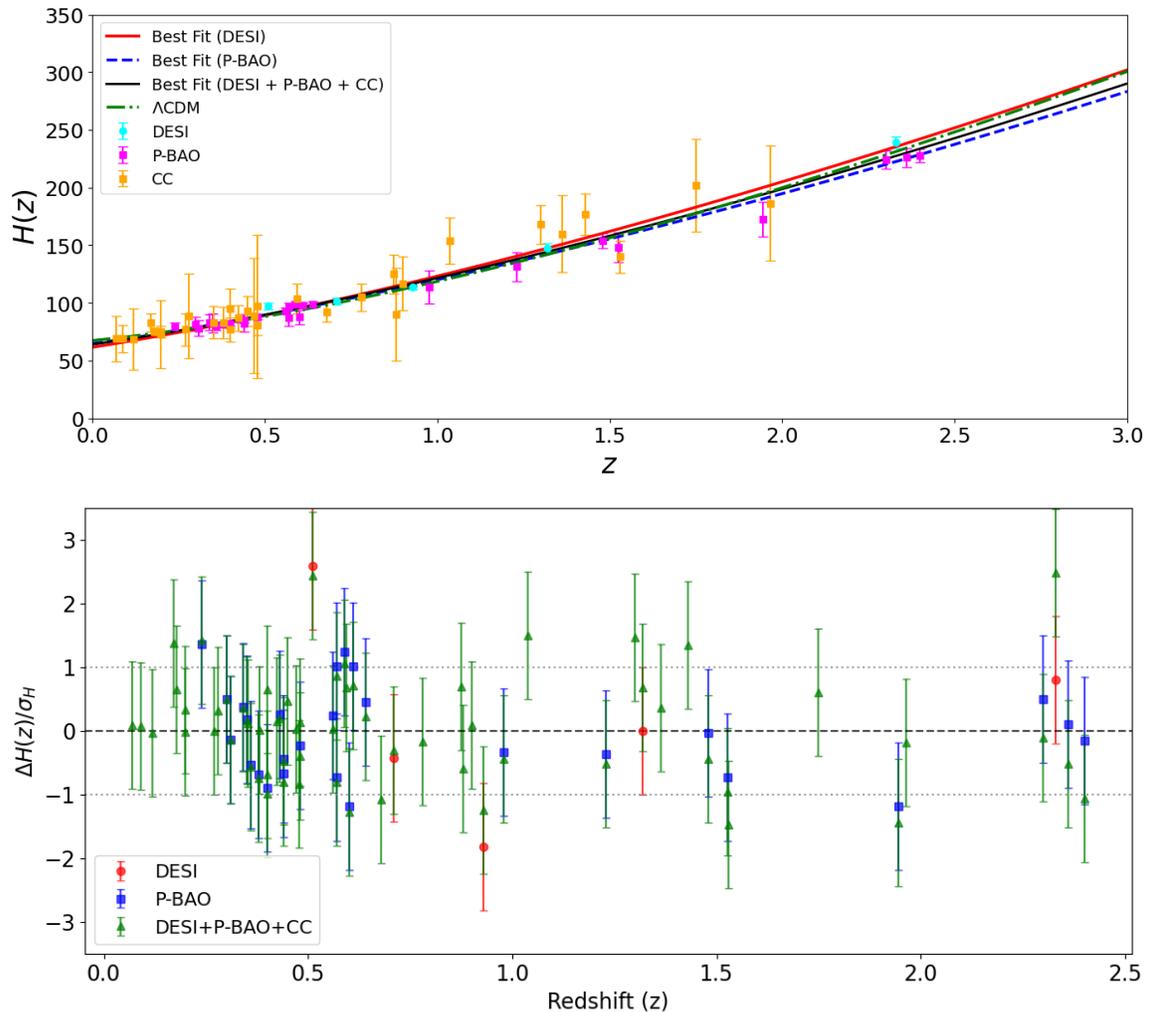

    \centering
    \includegraphics[width=0.85\textwidth]{p8_f1}\\[1ex]  
    \includegraphics[width=0.85\textwidth]{p8_f2}         
    \caption{The plot of $H(z)$ vs. $z$ for the model parameters that best suit the observational data is displayed in the upper panel. Additionally, a comparison with the $\Lambda$CDM model is shown. The normalised residual plot of $H(z)$ between the observational data and the model projection is displayed in the lower panel.}
    \label{f1}
\end{figure}

\begin{figure*}[htb]
\centerline{
\includegraphics[width=.8\textwidth]{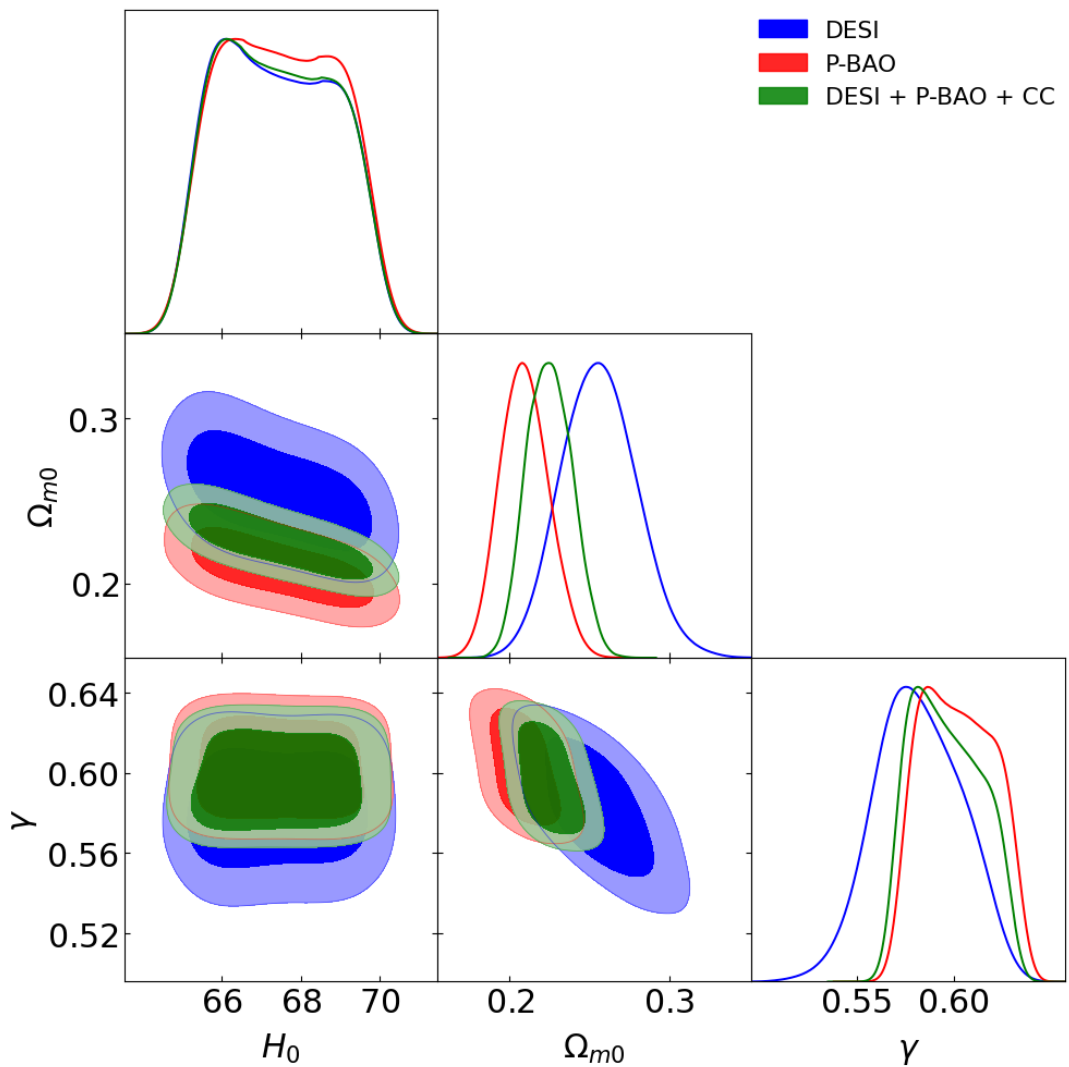}}
\caption{2-d contour sub-plot for the parameters $H_0$, $\omega_{m_0}$, and $\gamma$ with 1-$\sigma$ and 2-$\sigma$ errors (displaying the 68\% and 95\% c.l.) for $H(z)$ vs. $z$.}
\label{f2}
\end{figure*}

The constraints and cosmological consequences for the reconstructed $f(Q)$ gravity model are examined in this section using the observational datasets and technique outlined in Sec.~\ref{s4}. Tables~(\ref{tab1}) and (\ref{tabr1}) and Figs.~(\ref{f1}) and (\ref{f2}) display the results for which.
The performance of the reconstructed $f(Q)$ gravity model against observational data and the constraints on its parameters are shown in Figures~\ref{f1} and \ref{f2}. The best-fit evolution of the Hubble parameter $H(z)$ is compared with DESI, P-BAO, and combined datasets in the upper panel of Fig.~\ref{f1}, while the normalised residuals, which show excellent agreement with observations, are shown in the lower panel of Fig.~\ref{f1}. The 2D posterior contours for the vital model parameters ($H_0$, $\Omega_0$, $\gamma$) with $1\sigma$ and $2\sigma$ confidence regions are shown in Fig.~\ref{f2}, indicating that the parameters are tightly constrained. Here, table~(\ref{tab1}) summarises the best-fit values, which show $H_0 \sim 67.4$ km/s/Mpc, $\Omega_{m0} \sim 0.21$–$0.25$, and $\gamma \sim 0.58$–$0.60$ with small uncertainties. These values critically demonstrate the model's statistical robustness and reliability in explaining the observational data. For the present analysis using the relation $\gamma = \tfrac{3\alpha - 2}{2\alpha}$, one can infer the fractional parameter to be $\alpha \simeq 1.09$–$1.11$. This range lies comfortably within the theoretical bounds $1 < \alpha \leq 2$ derived in the foundational work of Oem Trivedi~\cite{Trivedi2024}, thereby confirming the internal consistency of our reconstructed FHDE-based $f(Q)$ model. Table~(\ref{tabr1}) results demonstrate that, when compared to the standard $\Lambda$CDM model, the reconstructed $f(Q)$ model consistently obtains higher $R^2$ values and lower $\chi^2_{\min}$, AIC, and BIC scores across all datasets. This improvement may be quantified by comparing the Akaike Information Criterion ($\Delta$AIC) and Bayesian Information Criterion ($\Delta$BIC) to standard $\Lambda$CDM. According to common interpretive ranges as mentioned in Sec.~\ref{s4}, we observe that for the DESI dataset $\Delta$AIC = 2.75 and $\Delta$BIC = 2.75, indicating positive evidence in favor of the $f(Q)$ model. Strong statistical support for the reconstructed $f(Q)$ model is also shown by the P-BAO dataset as it gives $\Delta$AIC = 8.90 and $\Delta$BIC = 8.90, as well as the combined DESI + P-BAO + CC datasets as it gives $\Delta$AIC = 6.02 and $\Delta$BIC = 6.02. These findings unequivocally reveal that the reconstructed $f(Q)$ model appears to be competitive in comparison to the $\Lambda$CDM model in terms of statistical robustness and offers a statistically better match to the data without the need for a cosmological constant.
Finally, we get the deceleration parameter, effective EoS, and $Om$ diagnostics for the model using Eq.~(\ref{e29c}) combined with Eqs.~(\ref{eq}), (\ref{ew}), and (\ref{eo}). Because of their complex structure, we do not write the explicit form of these quantities; instead, we present graphical representations for the model parameters' best-fit values, as shown in Figs.~\ref{f21}, (\ref{f22}), and (\ref{f23}). In addition, we assess the corresponding energy conditions for the reconstructed $f(Q)$ gravity model in order to demonstrate how well it complies with observational constraints. To further investigate the model's viability in terms of dynamical stability, we conduct a stability study of the rebuilt model under homogeneous linear perturbations.
\begin{table}[htbp]
\centering
\renewcommand{\arraystretch}{1.25}
\begin{tabular}{lcccc}
\toprule
\textbf{Dataset} & \boldmath$H_0$ & \boldmath$\Omega_{m0}$ & \boldmath$\gamma$ \\
\midrule
DESI & 
$67.43^{+1.70}_{-1.67}$ &
$0.2539^{+0.0247}_{-0.0225}$ &
$0.5826^{+0.0230}_{-0.0221}$ \\[4pt]
P-BAO &
$67.49^{+1.68}_{-1.68}$ &
$0.2085^{+0.0158}_{-0.0145}$ &
$0.6012^{+0.0221}_{-0.0193}$ \\[4pt]
DESI + P-BAO + CC &
$67.41^{+1.74}_{-1.65}$ &
$0.2246^{+0.0142}_{-0.0135}$ &
$0.5950^{+0.0225}_{-0.0181}$ \\
\bottomrule
\end{tabular}
\caption{Best-fit parameters with $1\sigma$ uncertainties for the reconstructed $f(Q)$ gravity model using DESI, P-BAO, and CC datasets.}
\label{tab1}
\end{table}

\begin{table*}[ht!] 
\centering
\renewcommand{\arraystretch}{1.25}
\begin{tabular}{llcccccc}
\toprule
\textbf{Model} & \textbf{Dataset} & \textbf{$R^2$} & \textbf{$\chi^2_{\min}$} & \textbf{AIC} & \textbf{BIC} & \textbf{$\Delta$AIC} & \textbf{$\Delta$BIC} \\
\midrule
{\textbf{New Model}} 
  & DESI               & 0.9937 & 10.83 & 16.83 & 15.66 & 2.75 & 2.75 \\
  & P-BAO              & 0.9892 & 12.89 & 18.89 & 22.78 & 8.90 & 8.90 \\
  & DESI + P-BAO + CC  & 0.9515 & 45.04 & 51.04 & 57.47 & 6.02 & 6.02 \\
\midrule
{\textbf{$\Lambda$CDM}} 
  & DESI               & 0.9878 & 13.58 & 19.58 & 18.41 & --   & --   \\
  & P-BAO              & 0.9829 & 21.79 & 27.79 & 31.68 & --   & --   \\
  & DESI + P-BAO + CC  & 0.9476 & 51.06 & 57.06 & 63.49 & --   & --   \\
\bottomrule
\end{tabular}
\caption{The reconstructed $f(Q)$ model and $\Lambda$CDM are statistically compared using DESI, P-BAO, and combined datasets. The formulas for $\Delta$AIC and $\Delta$BIC are $\mathrm{AIC}_{\Lambda\mathrm{CDM}}-\mathrm{AIC}_{\text{model}}$ and $\mathrm{BIC}_{\Lambda\mathrm{CDM}}-\mathrm{BIC}_{\text{model}}$, respectively. Preference for the $f(Q)$ model is shown by positive values.}
\label{tabr1}
\end{table*}

\begin{table}[htbp]
\centering
\renewcommand{\arraystretch}{1.25}
\begin{tabular}{lccc}
\toprule
\textbf{Dataset} & \boldmath$q(0)$ & \boldmath$\omega_{\mathrm{eff}}(0)$ & \boldmath$z_{\mathrm{tr}}$ \\
\midrule
DESI & $-0.3254^{+0.0684}_{-0.0638}$ & $-0.5503^{+0.0456}_{-0.0425}$ & $0.5577^{+0.0744}_{-0.0780}$ \\
P-BAO & $-0.4023^{+0.0547}_{-0.0488}$ & $-0.6016^{+0.0365}_{-0.0325}$ & $0.7227^{+0.0606}_{-0.0654}$ \\
DESI + P-BAO + CC & $-0.3736^{+0.0504}_{-0.0443}$ & $-0.5824^{+0.0336}_{-0.0295}$ & $0.6595^{+0.0527}_{-0.0578}$ \\
\bottomrule
\end{tabular}
\caption{Derived present-day cosmological indicators for the reconstructed $f(Q)$ gravity model, showing the deceleration parameter $q(0)$, effective equation-of-state parameter $\omega_{\mathrm{eff}}(0)$, and transition redshift $z_{\mathrm{tr}}$.}
\label{tab:newindicators}
\end{table}

\subsection{Deceleration Parameter}
\begin{figure*}[htb]
\centerline{
\includegraphics[width=1\textwidth]{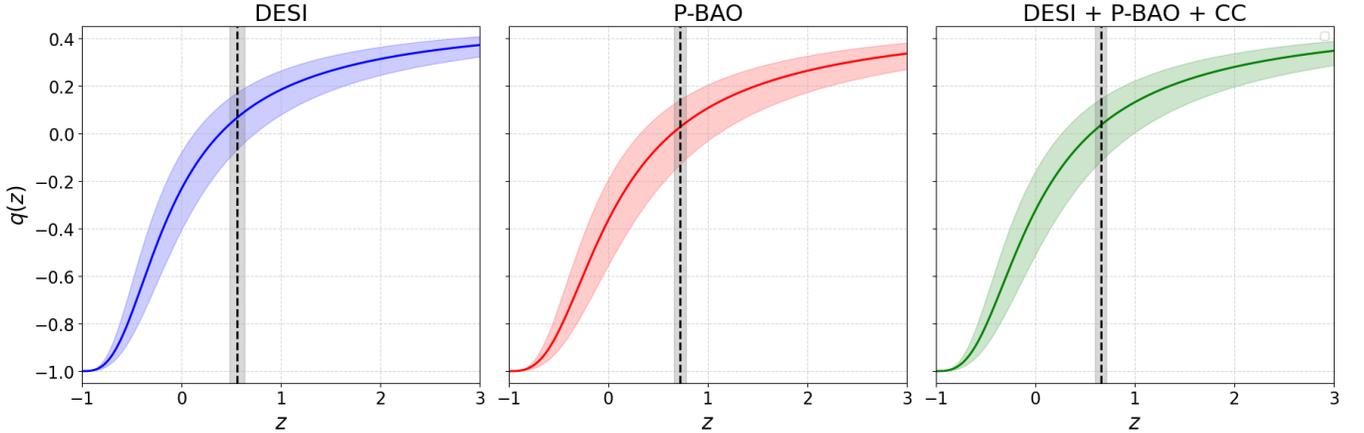}}
\caption{Deceleration parameter evolution with redshift and the corresponding transitional redshift for best model parameter values. The transitional redshift for each example is shown by the dotted line. Here, the allowed zone at a 1 $\sigma$ confidence level is shown by the shaded areas.}
\label{f21}
\end{figure*}
A key quantity in cosmology that describes the rate of change of the Universe's expansion is the deceleration parameter $q(z)$. It provides a direct measure of whether the cosmic expansion is accelerating or decelerating at a given epoch. A positive value of $q(z)$ corresponds to a decelerating Universe, as expected during the radiation- or matter-dominated eras, when the attractive nature of gravity slows down the expansion. In contrast, a negative value of $q(z)$ indicates an accelerated expansion, signaling the dominance of a repulsive component, such as dark energy or a modified gravitational effect, which counteracts the gravitational pull of matter and drives the late-time acceleration observed today. A variety of observations now firmly support the transition from deceleration to acceleration, which is thought to be a defining characteristic of late-time cosmic evolution.\\
The constraints from each dataset case (DESI, P-BAO, and the combined DESI + P-BAO + CC) for our reconstructed $f(Q)$ gravity model demonstrate that the evolution of $q(z)$ makes fine transitions from a decelerated to an accelerated phase. Also, it shows the present-day values of the deceleration parameter $q(0)$ to have a consistently negative $q(0)$ value, which provide confirmation for the current accelerated expansion of the Universe. Details illustrations for which are given by Fig. (\ref{f21}) and Table~\ref{tab:newindicators}. Results show $q(0) \in [-0.4023, -0.3254]$, which is compatible with independent observational estimates from Planck \cite{planck2018} and with recent studies addressing the Hubble tension \cite{verde2019tensions}. However, the magnitude of $q(0)$ suggests a moderate acceleration, with the P-BAO dataset showing slightly stronger acceleration compared to DESI. For all datasets, the precise transition redshift $z_{\text{tr}}$, at which the Universe transitions from deceleration to acceleration, falls within the range $z_{\rm tr} \in [0.5577, 0.7227]$, which is in good agreement with previous and recent studies that estimate the acceleration onset near $z \sim 0.6 - 0.8$ \cite{farooq2013, xu2012, xu2012new}. These results indicate that the reconstructed $f(Q)$ model successfully reproduces the observed late-time accelerated expansion of the Universe, achieving consistency with cosmological observations without the explicit introduction of a cosmological constant.

\subsection{Effective EoS}
\begin{figure*}[htb]
\centerline{
\includegraphics[width=1\textwidth]{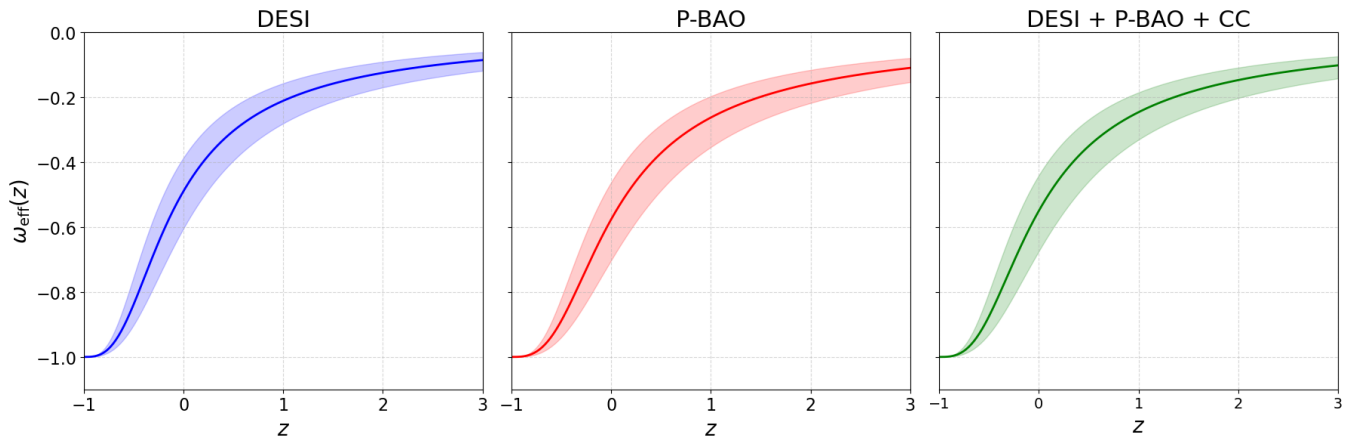}}
\caption{Evolution of effective EoS with redshift using best model parameter values for the reconstructed model. Here, the allowed zone at a 1 $\sigma$ confidence level is shown by the shaded areas.}
\label{f22}
\end{figure*}
The dynamical evolution of the cosmic fluid is fully described by the effective equation-of-state (EoS) parameter $\omega_{\rm eff}(z)$, which successfully incorporates the contributions from matter, radiation, and the geometrical modifications intrinsic to the reconstructed $f(Q)$ gravity model. Values of $\omega_{\rm eff}(z) < -1/3$ correspond to accelerated expansion, with $\omega_{\rm eff}(z) = -1$ representing the $\Lambda$CDM limit. If $\omega_{\rm eff}(z) < -1$, the Universe enters the phantom regime, whereas $-1 < \omega_{\rm eff}(z) \leq -1/3$ indicates a quintessence-like behavior.\\
From results the evolution of $\omega_{\rm eff}(z)$ for the reconstructed $f(Q)$ model shows a smooth transition from a matter-dominated decelerated phase at high redshift to an accelerated expansion at late times that resembles quintessence. The present-day values derived from the different datasets during the study consistently fall within the range $-0.6016 \lesssim \omega_{\rm eff}(0) \lesssim -0.5503$, suggesting that the Universe is currently undergoing an accelerated expansion that resembles quintessence. 
With respect to the present analysis the P-BAO dataset predicts a somewhat stronger acceleration, whereas the DESI dataset favours slightly less negative values, indicating milder acceleration.
The central estimate of $\omega_{\rm eff}(0) = -0.5824$ from the combined dataset (DESI + P-BAO + CC ) is consistent with current dynamical dark energy scenarios \cite{verde2019tensions} and in good agreement with observational constraints from Planck 2018 \cite{planck2018} as well. As shown in Fig.~\ref{f22}, the redshift evolution demonstrates that $\omega_{\rm eff}(z)$ decreases below $-1/3$ at low redshifts, driving cosmic acceleration, while at higher redshifts it approaches values near $0$, consistent with a matter-dominated decelerating Universe. Crucially, this transition happens naturally without the need for an explicit cosmological constant because of the dynamical contributions of the reconstructed $f(Q)$ component. Hence, these findings support the feasibility of the $f(Q)$ gravity model as a geometrical substitute for $\Lambda$CDM that is both consistent with early-time matter dominance and capable of reproducing the late-time acceleration. 
 
\subsection{$Om$ Diagnostics}
\begin{figure*}[htb]
\centerline{
\includegraphics[width=1\textwidth]{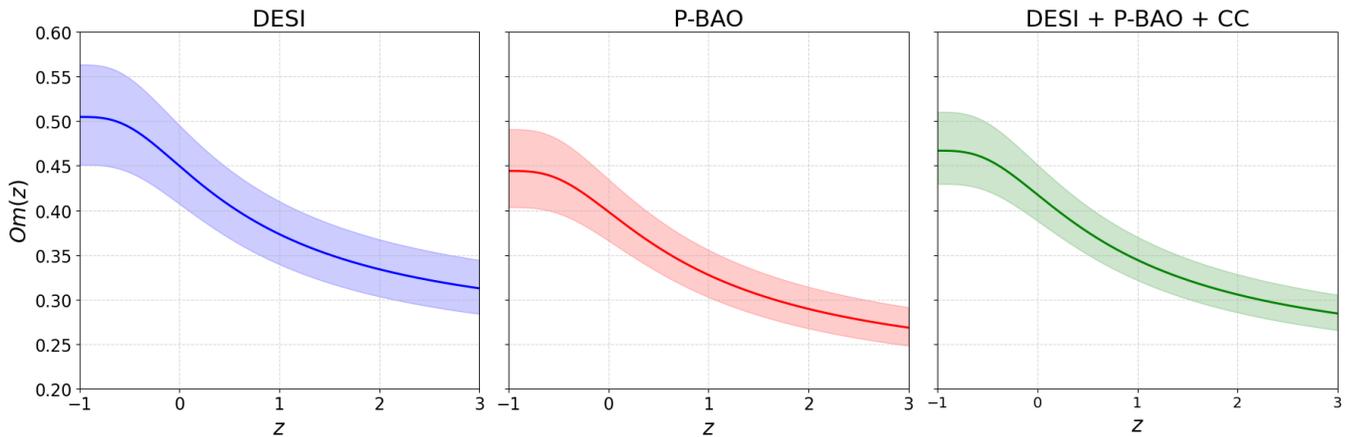}}
\caption{Evolution of $om$ diagnostic with redshift using best model parameter values for the reconstructed model. Here, the allowed zone at a 1 $\sigma$ confidence level is shown by the shaded areas.}
\label{f23}
\end{figure*}
Without assuming any particular form for the equation of state (EoS), the $Om(z)$ diagnostic is a reliable, geometry-based tool for characterising the Universe's expansion history and differentiating between competing dark energy models. With respect to the standard $\Lambda$CDM model, $Om(z)$ is expected to remain constant across redshifts. And, any kind of departures or deviations from this constant behaviour can hint at alternative dark energy dynamics, as mentioned below:
\begin{itemize}
    \item \textbf{Constant} $Om(z)$: $\Lambda$CDM-like behavior.
    \item \textbf{Negative slope}: Quintessence-like behavior ($\omega > -1$).
    \item \textbf{Positive slope}: Phantom-like behavior ($\omega < -1$).
\end{itemize}
As shown in Fig.~\ref{f23}, the $Om(z)$ diagnostic for our reconstructed $f(Q)$ gravity model clearly shows a negative slope across all datasets involved in the study. This systematic decline in $Om(z)$ with respect to redshift is especially evident in the low-$z$ regime, where deviations from the $\Lambda$CDM constant line become more noticeable. Such a trend indicates that the reconstructed $f(Q)$ gravity model favours a quintessence-like dark energy behavior, characterized by an effective equation of state parameter satisfying the condition $\omega_{\rm eff} > -1$. The results drawn from the deceleration parameter $q(z)$ and the effective EoS $\omega_{\rm eff}(z)$ analyses agree very well with the observed negative gradient of $Om(z)$. When taken as a whole, the results of these diagnostics consistently indicate that a dynamical, quintessence-type component, rather than a static cosmological constant, is driving the late-time accelerating phase. This precise agreement among independent geometrical probes strengthens the reliability of the reconstructed $f(Q)$ gravity model as a viable alternative framework for explaining dark energy, one that can naturally account for the observed accelerated expansion of the universe, through geometric modifications of gravity rather than through the ad hoc inclusion of a constant vacuum energy term.
\subsection{Energy Conditions}
We investigate how the classical energy conditions behave as functions of redshift $z$ in order to further evaluate the physical credibility of the reconstructed $f(Q)$ gravity model. These energy conditions act as important consistency check points, ensuring that the effective energy density and pressure derived from the modified field equations satisfy physically reasonable bounds. They are expressed in terms of the effective energy density $\rho_{\text{eff}}$ and pressure $p_{\text{eff}}$ as given below:
\begin{itemize}
    \item \textbf{Weak Energy Condition (WEC)}: $\rho_{\text{eff}} \geq 0$, \hspace{3mm} $\rho_{\text{eff}} + p_{\text{eff}} \geq 0$
    \item \textbf{Null Energy Condition (NEC)}: \hspace{6mm} $\rho_{\text{eff}} + p_{\text{eff}} \geq 0$
    \item \textbf{Strong Energy Condition (SEC)}: \hspace{3mm} $\rho_{\text{eff}} + 3p_{\text{eff}} \geq 0$
    \item \textbf{Dominant Energy Condition (DEC)}: \hspace{2mm} $\rho_{\text{eff}} - p_{\text{eff}} \geq 0$
\end{itemize}

In this framework, $\rho_{\text{eff}}$ and $p_{\text{eff}}$ are determined from the modified Friedmann equations of the given $f(Q)$ cosmological model:
\begin{equation}
3H^2 = \rho_{\text{eff}},
\label{q1}
\end{equation}
\begin{equation}
2\dot{H} + 3H^2 = -p_{\text{eff}}.
\label{q2}
\end{equation}
where $H$ denotes the Hubble parameter. These relations allow us to investigate how the effective fluid derived from the reconstructed geometry evolves with redshift and whether it respects or violates the standard energy conditions across cosmic time.
\begin{figure*}[htb]
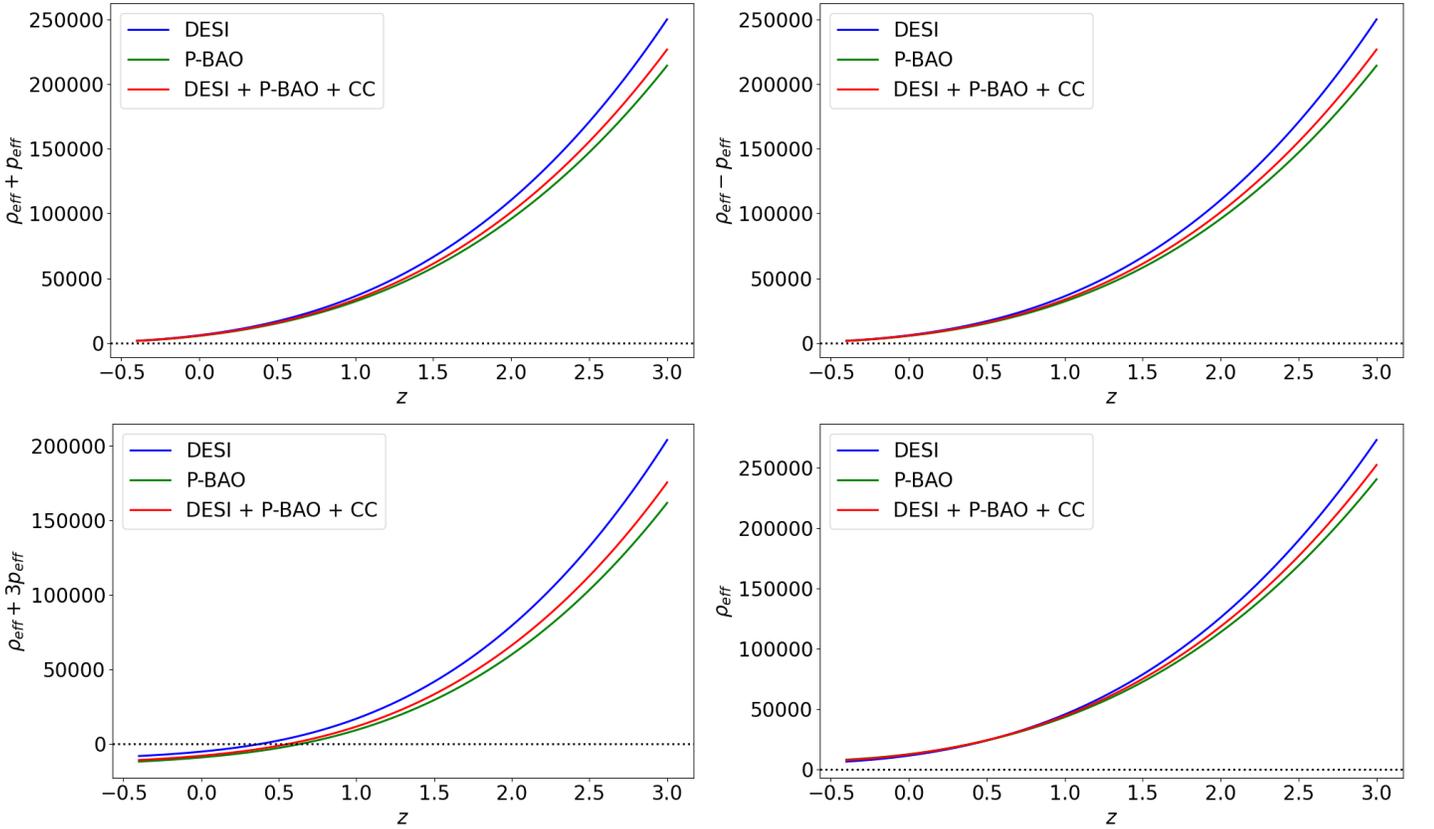

\centerline{
\includegraphics[width=.52\textwidth]{SEC}
\includegraphics[width=.52\textwidth]{NEC}}
\centerline{
\includegraphics[width=.52\textwidth]{DEC}
\includegraphics[width=.52\textwidth]{WEC}}
\caption{Redshift evolution of the energy conditions for best model parameter values.}
\label{fe}
\end{figure*}
Using the best-fit cosmological parameters obtained from the MCMC analysis based on the DESI, P-BAO and CC datasets, we examine the redshift evolution of the classical energy conditions by employing Eq.~(\ref{e29c}) together with Eqs.~(\ref{q1}) and (\ref{q2}). The corresponding results are presented in Fig.~(\ref{fe}). From the figure, it is evident that both the Weak Energy Condition (WEC) and the Dominant Energy Condition (DEC) remain satisfied throughout the entire cosmological history. This ensures the positivity of the effective energy density and confirms that the effective pressure does not exceed the energy density, thereby indicating a physically admissible, non-exotic cosmic fluid. Additionally, the reconstructed $f(Q)$ gravity model maintains the Null Energy Condition (NEC) across all the all redshifts, which is frequently broken in phantom like scenarios. This preservation indicates the absence of phantom instabilities and presence of a smooth and dynamically stable dark energy behaviour. On the other hand, the Strong Energy Condition (SEC) is found to be violated at low redshifts ($z \lesssim 1$), which is consistent with the beginning of the current accelerated expansion of the Universe. However, at higher redshifts ($z \gg 1$) the SEC is restored, pointing out the expected transition from a decelerating, matter-dominated epoch to the current phase of cosmic acceleration.\\
Overall, these results show that without the need for a cosmological constant or the use of phantom energy, the reconstructed $f(Q)$ gravity model naturally bridges the decelerated and accelerated eras of cosmic evolution. This highlights the model's ability to reproduce the late-time acceleration as a purely geometric effect, as well as its physical consistency and observational compatibility.
\subsection{Stability Analysis}
\begin{figure*}[htb]
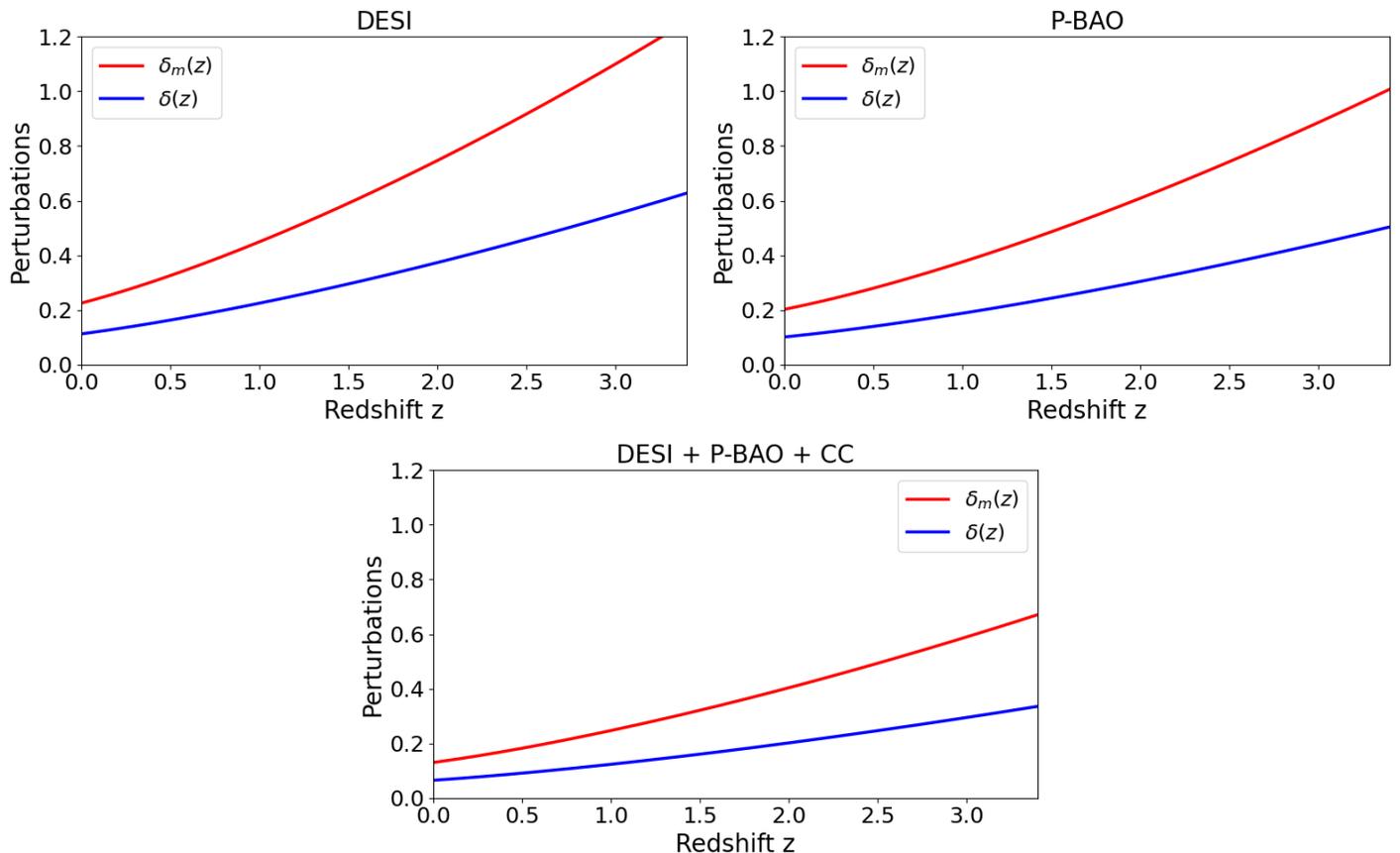

\centerline{
\includegraphics[width=.52\textwidth]{p_1}
\includegraphics[width=.52\textwidth]{p_2}}
\centerline{
\includegraphics[width=.52\textwidth]{p_3}}
\caption{The evolution of the perturbation terms $\delta_m(z)$ and $\delta(z)$ for each dataset's best model parameters as a function of redshift $z$.}
\label{8f}
\end{figure*}

The reconstructed cosmological model's dynamical stability against homogeneous linear perturbations is investigated in this section. To achieve so, we introduce first-order perturbations in the energy density and the Hubble parameter \cite{ref90AN,ref90AN1}, which are represented as shown below:
\begin{equation}
H^{*}(t) = H(t)\,[1 + \delta(t)],
\label{An1}
\end{equation}
\begin{equation}
\rho^{*}(t) = \rho(t)\,[1 + \delta_m(t)],
\label{An2}
\end{equation}
where $H^{*}(t)$ and $\rho^{*}(t)$ represents the perturbed Hubble parameter and energy density, while $\delta(t)$ and $\delta_m(t)$ represent their respective fractional perturbations.\\
Literature shows in standard cosmology, the conservation of energy–momentum is governed by the continuity equation,
\begin{equation}
\dot{\rho} + 3H(\rho + p) = 0,
\label{An3}
\end{equation}
where the overdot represents differentiation with respect to cosmic time $t$. Also,here the time derivative can be conveniently related to the redshift $z$ through:
\begin{equation}
\frac{d}{dt} = - (1 + z)\, H(z)\, \frac{d}{dz}.
\label{r1}
\end{equation}
Substituting Eqs.~(\ref{An1}), (\ref{An2}), (\ref{ew}), and (\ref{q1}) into Eq.~(\ref{An3}) yields the following first-order differential relations:
\begin{equation}
\dot{\delta}_m(t) + 3H\,(1 + \omega_{\rm eff})\, \delta(t) = 0,
\label{An4}
\end{equation}
\begin{equation}
2\,\delta(t) = \delta_m(t).
\label{An5}
\end{equation}
Using Eq.~(\ref{r1}) and transforming the above relations into redshift space, we obtain:
\begin{equation}
-(1 + z)\, \frac{d\delta_m}{dz} + \frac{2}{3}\left(1 + \omega_{\rm eff}(z)\right)\delta_m = 0,
\label{An4a}
\end{equation}
\begin{equation}
2\,\delta(z) = \delta_m(z),
\label{An5a}
\end{equation}
where the effective equation of state parameter is given by
\begin{equation}
\omega_{\rm eff}(z) = -1 + \frac{2(1 + z)}{3H(z)} \frac{dH}{dz}.
\end{equation}
Here, we can trace the redshift evolution of the perturbation terms $\delta(z)$ and $\delta_m(z)$ by solving Eq.~\ref{An4a}. Using the best-fit parameter values of the reconstructed model under different observational datasets, the resulting behaviour of these perturbation terms is shown in Fig.~\ref{8f}. These results shed light on the model's stability features and compatibility with cosmic dynamics that are supported by observations.\\
The evolution of the perturbation functions $\delta_m(z)$ and $\delta(z)$ reveals a distinctive monotonic decrease with the decline of redshift $z$, as shown in Fig.~\ref{8f}. This would indicate that both perturbations decrease as the Universe evolves towards its current epoch. Such smooth decay in perturbation amplitudes suggests that there is no late-time growth of instabilities and that the reconstructed model is dynamically stable under linear homogeneous perturbations. Conclusively, this uniform behavior of the damping trend for the perturbations across all observational datasets provides strong evidence for the internal consistency and physical robustness of the reconstructed model, thereby reinforcing its suitability as a viable alternative approach for explaining the presently observed cosmic accelerated expansion of the Universe.
\section{Conclusion}\label{c1}
In this work, we have presented a novel reconstruction of the $f(Q)$ gravity framework inspired by the Fractional Holographic Dark Energy (FHDE) scenario, aiming to provide a geometrically motivated explanation for the late-time acceleration of the Universe without invoking an explicit cosmological constant. By incorporating the fractional holographic dark energy formalism and adopting the Hubble horizon as the infrared cutoff, we obtained an analytical form of the reconstructed $f(Q)$ gravity function, given as $f(Q)=Q+F(Q), \quad F(Q) = \frac{6^{1-\gamma } c^2 Q^{\gamma } \left(1-2 \gamma  \left(\frac{Q}{Q_0}\right)^{\frac{1}{2}-\gamma }\right)}{1-2 \gamma }$. This reconstruction ensures that General Relativity is recovered in the appropriate limit ($c \to 0$) while introducing a rich cosmological dynamics compatible with current observational evidence.\\
The free parameters of the reconstructed model were constrained using the latest DESI BAO data, a previous BAO compilation (P-BAO), and cosmic chronometer (CC) datasets through a detailed Markov Chain Monte Carlo (MCMC) analysis. Statistical diagnostics, including $R^2$, $\chi^2_{\min}$, AIC, and BIC, consistently show that the reconstructed $f(Q)$ model showed a comparative and effective performance than the standard $\Lambda$CDM cosmology across all datasets, providing strong evidence in favor of this geometrically modified gravity approach. We also examined the cosmological implications proposed by the reconstructed model based on the best fit values. Our analysis shows that, at low redshifts, the model effectively replicates the observed cosmic transition from a decelerated, matter-dominated era to an accelerated expansion phase. For all the datasets invloved in the study, the present-day value of the deceleration parameter $q(0)$ lies in the negative, confirming the ongoing accelerated expansion of the universe. The reconstructed $f(Q)$ gravity model's non-$\Lambda$CDM, dynamical dark energy nature is further supported by the $Om(z)$ diagnostic's mild negative slope and the effective equation-of-state parameter $\omega_{\mathrm{eff}}(z)$ staying within the range $-1 < \omega_{\mathrm{eff}}(0) < -1/3$, which corresponds to a regime that is quintessence-like. The physical admissibility of the effective fluid represented by the model is guaranteed by the analysis of classical energy conditions, which shows that the Weak and Dominant Energy Conditions are met throughout the cosmic evolution. The Null Energy Condition, which is often seen to violated in phantom models, remains valid for the current model, signifying the absence of phantom like behavior and associated instabilities. The observed late-time acceleration is consistent with the violation of the Strong Energy Condition at low redshifts ($z \lesssim 1$), but the expected matter-dominated epoch is reflected at higher redshifts ($z \gg 1$). In order to assess the dynamical stability of the model,  we also investigated linear homogeneous perturbations in the Hubble parameter and the energy density for the model. Where we see the perturbation modes $\delta(z)$ and $\delta_m(z)$ exhibit a monotonic decay with decreasing redshift, suggesting that the model does not develop late-time instabilities and is stable under small perturbations. The reconstructed framework's internal consistency and dynamical soundness are confirmed by this robustness across several datasets.\\
All things taken into consideration, the reconstructed $f(Q)$ model offers a coherent and empirically validated account of the cosmic evolution of the universe. Without the need for exotic matter fields or a cosmological constant, it naturally explains the change from a decelerating to an accelerating phase. The reconstructed model thus represents a geometrically motivated, stable, and physically viable substitute to $\Lambda$CDM, capable of describing the late-time accelerated expansion of the universe, only by changing the fundamental gravitational geometry. To further confirm its cosmological relevance, future extensions of the present work might examine structure formation, gravitational wave propagation, and CMB constraints within this reconstructed $f(Q)$gravity model. Also, an immediate extension of this work would be to repeat the reconstruction using alternative infrared cutoffs, for instance the event horizon or the particle horizon, in place of the Hubble horizon. It is also worthwhile to consider more involved cutoffs such as the Granda–Oliveros or generalized Nojiri–Odintsov cutoffs, to assess the robustness of the reconstructed $f(Q)$ an the cosmological dynamics it forecasts.


\begin{thebibliography}{200}





\bibitem{Riess1998}
A.~G.~Riess et al., ``Observational evidence from supernovae for an accelerating universe and a cosmological constant,'' \textit{Astron. J.} \textbf{116}, 1009 (1998), \href{https://arxiv.org/abs/astro-ph/9805201}{arXiv:astro-ph/9805201}.

\bibitem{Perlmutter1999}
S.~Perlmutter et al., ``Measurements of Omega and Lambda from 42 high-redshift supernovae,'' \textit{Astrophys. J.} \textbf{517}, 565 (1999), \href{https://arxiv.org/abs/astro-ph/9812133}{arXiv:astro-ph/9812133}.

\bibitem{Astier2006}
P.~Astier et al., ``The Supernova Legacy Survey: Measurement of $\Omega_M$, $\Omega_Lambda$ and $\omega$ from the first year data set,'' \textit{Astron. Astrophys.} \textbf{447}, 31 (2006), \href{https://arxiv.org/abs/astro-ph/0510447}{arXiv:astro-ph/0510447}.

\bibitem{Spergel2003}
D.~N.~Spergel et al., ``First-Year Wilkinson Microwave Anisotropy Probe (WMAP) Observations: Determination of Cosmological Parameters,'' \textit{Astrophys. J. Suppl. Ser.} \textbf{148}, 175 (2003), \href{https://arxiv.org/abs/astro-ph/0302209}{arXiv:astro-ph/0302209}.

\bibitem{Tegmark2004}
M.~Tegmark et al., ``Cosmological parameters from SDSS and WMAP,'' \textit{Phys. Rev. D} \textbf{69}, 103501 (2004), \href{https://arxiv.org/abs/astro-ph/0310723}{arXiv:astro-ph/0310723}.

\bibitem{Cole2005}
S.~Cole et al., ``The 2dF Galaxy Redshift Survey: power-spectrum analysis of the final dataset and cosmological implications,'' \textit{Mon. Not. R. Astron. Soc.} \textbf{362}, 505 (2005), \href{https://arxiv.org/abs/astro-ph/0501174}{arXiv:astro-ph/0501174}.

\bibitem{Eisenstein2005}
D.~J.~Eisenstein et al., ``Detection of the Baryon Acoustic Peak in the Large-Scale Correlation Function of SDSS Luminous Red Galaxies,'' \textit{Astrophys. J.} \textbf{633}, 560 (2005), \href{https://arxiv.org/abs/astro-ph/0501171}{arXiv:astro-ph/0501171}.

\bibitem{Weinberg1989}
S.~Weinberg, ``The cosmological constant problem,'' \textit{Rev. Mod. Phys.} \textbf{61}, 1 (1989).

\bibitem{Martin2012}
J.~Martin, ``Everything You Always Wanted To Know About The Cosmological Constant Problem (But Were Afraid To Ask),'' \textit{Comptes Rendus Physique} \textbf{13}, 566 (2012), \href{https://arxiv.org/abs/1205.3365}{arXiv:1205.3365 [astro-ph.CO]}.


\bibitem{za1}
Yang, Y., Ren, X., Wang, Q., Lu, Z., Zhang, D., Cai, Y.-F., \& Saridakis, E. N. (2024). 
Quintom cosmology and modified gravity after DESI 2024. 
\emph{Science Bulletin}, 69(17), 2698–2704. 
\href{https://doi.org/10.1016/j.scib.2024.07.029}{https://doi.org/10.1016/j.scib.2024.07.029}

\bibitem{za2}
Yang, Y., Ren, X., Wang, B., Cai, Y.-F., \& Saridakis, E. N. (2024). 
Data reconstruction of the dynamical connection function in $f(Q)$ cosmology. 
\emph{Monthly Notices of the Royal Astronomical Society}, 533(2), 2232–2241. 
\href{https://doi.org/10.1093/mnras/stae1905}{https://doi.org/10.1093/mnras/stae1905}

\bibitem{za3}
Paliathanasis, A. (2025). 
Testing Non-Coincident $f(Q)$-gravity with DESI DR2 BAO and GRBs. 
\emph{arXiv preprint} \href{https://arxiv.org/abs/2504.11132}{arXiv:2504.11132}.

\bibitem{za4}
Paliathanasis, A. (2023). 
Dynamical analysis of $f(Q)$-cosmology. 
\emph{arXiv preprint} \href{https://arxiv.org/abs/2304.04219}{arXiv:2304.04219}.

\bibitem{za5}
Dimakis, N., Roumeliotis, M., Paliathanasis, A., Apostolopoulos, P. S., \& Christodoulakis, T. (2022). 
Self-similar cosmological solutions in symmetric teleparallel theory: FLRW spacetimes. 
\emph{Phys. Rev. D}, \textbf{106}(12), 123516. 
\href{https://doi.org/10.1103/PhysRevD.106.123516}{doi:10.1103/PhysRevD.106.123516}

\bibitem{za6}
Basilakos, S., Paliathanasis, A., \& Saridakis, E. N. (2025). 
Equivalence of $f(Q)$ cosmology with quintom-like scenario: The phantom field as effective realization of the non-trivial connection. 
\emph{Phys. Lett. B}, \textbf{868}, 139658. 
\href{https://doi.org/10.1016/j.physletb.2025.139658}{doi:10.1016/j.physletb.2025.139658}

\bibitem{za7}
Khyllep, W., Paliathanasis, A., \& Dutta, J. (2021). 
Cosmological solutions and growth index of matter perturbations in $f(Q)$ gravity. 
\emph{Phys. Rev. D}, \textbf{103}(10), 103521. 
\href{https://doi.org/10.1103/PhysRevD.103.103521}{doi:10.1103/PhysRevD.103.103521}

\bibitem{za8}
Nojiri, S., \& Odintsov, S. D. (2011). 
Unified cosmic history in modified gravity: From $F(R)$ theory to Lorentz non-invariant models. 
\emph{Phys. Rept.}, \textbf{505}, 59–144. 
\href{https://doi.org/10.1016/j.physrep.2011.04.001}{doi:10.1016/j.physrep.2011.04.001}

\bibitem{za9}
Nojiri, S., \& Odintsov, S. D. (2024). 
Well-defined $f(Q)$ gravity, reconstruction of FLRW spacetime and unification of inflation with dark energy epoch. 
\emph{Phys. Dark Univ.}, \textbf{45}, 101538. 
\href{https://doi.org/10.1016/j.dark.2024.101538}{doi:10.1016/j.dark.2024.101538}

\bibitem{za10}
Nojiri, S., \& Odintsov, S. D. (2024). 
$f(Q)$ Gravity with Gauss–Bonnet Corrections: From Early-Time Inflation to Late-Time Acceleration. 
\emph{Fortschritte der Physik}, \textbf{72}(9-10), 2400113. 
\href{https://doi.org/10.1002/prop.202400113}{doi:10.1002/prop.202400113}



\bibitem{Cai2007}
R.~G.~Cai, ``A Dark Energy Model Characterized by the Age of the Universe,'' \textit{Phys. Lett. B} \textbf{657}, 228 (2007), \href{https://arxiv.org/abs/0707.4049}{arXiv:0707.4049 [hep-th]}.

\bibitem{Wei2007}
H.~Wei and R.~G.~Cai, ``A New Model of Agegraphic Dark Energy,'' \textit{Phys. Lett. B} \textbf{660}, 113 (2008), \href{https://arxiv.org/abs/0708.0884}{arXiv:0708.0884 [astro-ph]}.


\bibitem{Trivedi2024}
Trivedi, O., Bidlan, A., \& Moniz, P. (2024). Fractional holographic dark energy. \emph{Physics Letters B}, \textbf{858}, 139074--139074. 
\href{https://doi.org/10.1016/j.physletb.2024.139074}{doi:10.1016/j.physletb.2024.139074}

\bibitem{Rathore2025}
Rathore, S., Meetei, E. C., \& Surendra, S. S. (2025). Observational viability of fractional holographic dark energy in LRS Bianchi type-I cosmological model. \emph{arXiv preprint}. \href{https://arxiv.org/abs/2506.16004}{https://arxiv.org/abs/2506.16004}

\bibitem{Bidlan2025}
Bidlan, A., Moniz, P., \& Trivedi, O. (2025). Reconstructing Fractional Holographic Dark Energy with scalar and gauge fields. \emph{Eur. Phys. J. C}, \textbf{85}, 520. 
\href{https://doi.org/10.1140/epjc/s10052-025-14238-2}{doi:10.1140/epjc/s10052-025-14238-2}

\bibitem{Capozziello2006}
S.~Capozziello, V.~F.~Cardone, and A.~Troisi, ``Dark energy and dark matter as curvature effects,'' \textit{J. Cosmol. Astropart. Phys} \textbf{0608}, 001 (2006), \href{https://arxiv.org/abs/astro-ph/0602349}{arXiv:astro-ph/0602349}.

\bibitem{Nojiri2006recon}
S.~Nojiri and S.~D.~Odintsov, ``Modified f(R) gravity consistent with realistic cosmology: From matter dominated epoch to dark energy universe,'' \textit{Phys. Rev. D} \textbf{74}, 086005 (2006), \href{https://arxiv.org/abs/hep-th/0608008}{arXiv:hep-th/0608008}.

\bibitem{Nojiri2011rev}
Nojiri, S., \& Odintsov, S. D. (2011).
Unified cosmic history in modified gravity: from $F(R)$ theory to Lorentz non-invariant models.
\emph{Physics Reports}, 505(2–4), 59–144.
\href{https://doi.org/10.1016/j.physrep.2011.04.001}{https://doi.org/10.1016/j.physrep.2011.04.001}



\bibitem{Wu2010}
Wu, P., \& Yu, H. (2010).
Reconstruction of the deceleration parameter and the equation of state of dark energy.
\emph{Physics Letters B}, 693(4), 415–420.
\href{https://doi.org/10.1016/j.physletb.2010.08.031}{https://doi.org/10.1016/j.physletb.2010.08.031}

\bibitem{Bamba2011recon}
Bamba, K., Capozziello, S., Nojiri, S., \& Odintsov, S. D. (2012).
Dark energy cosmology: the equivalent description via different theoretical models and cosmography tests.
\emph{Astrophysics and Space Science}, 342(1), 155–228.
\href{https://doi.org/10.1007/s10509-012-1181-8}{https://doi.org/10.1007/s10509-012-1181-8}


\bibitem{Harko2018}
Harko, T., Koivisto, T. S., Lobo, F. S. N., Olmo, G. J., \& Rubiera-Garcia, D. (2018).
Coupling matter in modified $Q$ gravity.
\emph{Physical Review D}, 98(8), 084043.
\href{https://doi.org/10.1103/PhysRevD.98.084043}{https://doi.org/10.1103/PhysRevD.98.084043}

\bibitem{Sotiriou2010}
T.~P.~Sotiriou and V.~Faraoni, ``$f(R)$ theories of gravity,'' \textit{Rev. Mod. Phys.} \textbf{82}, 451 (2010), \href{https://arxiv.org/abs/0805.1726}{arXiv:0805.1726 [gr-qc]}.

\bibitem{Jimenez2018}
J.~B.~Jiménez, L.~Heisenberg, and T.~Koivisto, ``Coincident General Relativity,'' \textit{Phys. Rev. D} \textbf{98}, 044048 (2018), \href{https://arxiv.org/abs/1710.03116}{arXiv:1710.03116 [gr-qc]}.

\bibitem{Mandal2020}
S.~Mandal, P.~K.~Sahoo, and J.~R.~L.~Santos, ``Cosmography in f(Q) gravity,'' \textit{Phys. Rev. D} \textbf{102}, 024057 (2020), \href{https://arxiv.org/abs/2001.02357}{arXiv:2001.02357 [gr-qc]}.

\bibitem{Dunsby2010}
P.~K.~S.~Dunsby, E.~Elizalde, R.~Goswami, S.~Odintsov, and D.~S.~Gomez, ``On the LCDM Universe in f(R) gravity,'' \textit{Phys. Rev. D} \textbf{82}, 023519 (2010), \href{https://arxiv.org/abs/1005.2205}{arXiv:1005.2205 [gr-qc]}.

\bibitem{Goheer2009}
N.~Goheer, R.~Goswami, and P.~K.~S.~Dunsby, ``Dynamics of $f(G)$ cosmology,'' \textit{Phys. Rev. D} \textbf{79}, 121301 (2009), \href{https://arxiv.org/abs/0904.2559}{arXiv:0904.2559 [gr-qc]}.

\bibitem{Baffou2017}
E.~H.~Baffou, M.~J.~S.~Houndjo, and J.~Tossa, ``Reconstruction of $f(\tau,T)$ gravity in different cosmological models,'' \textit{Eur. Phys. J. C} \textbf{77}, 708 (2017), \href{https://arxiv.org/abs/1706.08842}{arXiv:1706.08842 [gr-qc]}.

\bibitem{Saha2025}
P.~Saha and P.~Rudra, ``A Cosmological Holographic Reconstruction of $f(Q)$ Theory,'' \textit{arXiv:2407.01870v2} (2025), \href{https://arxiv.org/abs/2407.01870}{arXiv:2407.01870 [gr-qc]}.

\bibitem{f1}
 S. Nojiri and S.D. Odintsov, Introduction to modified gravity and gravitational alternative
 for dark energy, Int. J. Geom. Meth. Mod. Phys. 4 (2007) 115-145.\href{https://doi.org/10.1142/S0219887807001928}{doi:10.1142/S0219887807001928}
 
\bibitem{Karami2010}
Xiang-Lai Liu \& Xin Zhang, ``New agegraphic dark energy in Brans-Dicke theory,'' \textit{Commun. Theor. Phys.}, \textbf{52}, 761 (2009), \href{https://arxiv.org/abs/0909.4911}{arXiv:0909.4911 [astro-ph.CO]}.

\bibitem{Karami2011a}
K.~Karami and M.~S.~Khaledian, ``Reconstructing $f(R)$ modified gravity from entropy-corrected versions of holographic and new agegraphic dark energy models,'' \textit{J. High Energy Phys.}, \textbf{2011}, 086 (2011), \href{https://arxiv.org/abs/1004.1805}{arXiv:1004.1805 [gr-qc]}.

\bibitem{Granda2008}
L. N. Granda and A. Oliveros, ``Infrared cut-off proposal for the holographic density,'' \textit{Phys. Lett. B} \textbf{669}, 275–277 (2008), \href{https://doi.org/10.1016/j.physletb.2008.10.017}{doi:10.1016/j.physletb.2008.10.017}.

\bibitem{Kim2013}
H.-C. Kim, J.-W. Lee, and J. Lee, ``Causality problem in a holographic-dark-energy model,'' \textit{EPL (Europhys. Lett.)} \textbf{102}, 29001 (2013), \href{https://doi.org/10.1209/0295-5075/102/29001}{doi:10.1209/0295-5075/102/29001}.

\bibitem{Nojiri2017}
S. Nojiri and S. D. Odintsov, ``Covariant generalized holographic dark energy and accelerating universe,'' \textit{Eur. Phys. J. C} \textbf{77}, 528 (2017), \href{https://doi.org/10.1140/epjc/s10052-017-5097-x}{doi:10.1140/epjc/s10052-017-5097-x}.

\bibitem{Nojiri2019}
S. Nojiri, S. D. Odintsov, and E. N. Saridakis, ``Holographic bounce,'' \textit{Nucl. Phys. B} \textbf{949}, 114790 (2019), \href{https://doi.org/10.1016/j.nuclphysb.2019.114790}{doi:10.1016/j.nuclphysb.2019.114790}.

\bibitem{Li2004}
M. Li, ``A model of holographic dark energy,'' \textit{Phys. Lett. B} \textbf{603}, 1–5 (2004), \href{https://doi.org/10.1016/j.physletb.2004.10.014}{doi:10.1016/j.physletb.2004.10.014}.

\bibitem{Myung2007}
Y. S. Myung, ``Instability of holographic dark energy models,'' \textit{Phys. Lett. B} \textbf{652}, 223–227 (2007), \href{https://doi.org/10.1016/j.physletb.2007.07.033}{doi:10.1016/j.physletb.2007.07.033}.

\bibitem{ref98} Adame, A. G., et al. (2024). DESI 2024 VI: Cosmological Constraints from the Measurements of Baryon Acoustic Oscillations. \textit{arXiv}. \href{https://arxiv.org/abs/2404.03002}{https://arxiv.org/abs/2404.03002}

\bibitem{d113} Gaztanaga, E., Cabre, A., \& Hui, L. (2009). Clustering of Luminous Red Galaxies IV: Baryon Acoustic Peak in the Line-of-Sight Direction and a Direct Measurement of H(z). \textit{Monthly Notices of the Royal Astronomical Society}, \textit{399}, 1663–1680. \href{https://arxiv.org/abs/0807.3551}{arXiv:0807.3551}

\bibitem{d114} Oka, A., Saito, S., Nishimichi, T., Taruya, A., \& Yamamoto, K. (2014). Simultaneous constraints on the growth of structure and cosmic expansion from the multipole power spectra of the SDSS DR7 LRG sample. \textit{Monthly Notices of the Royal Astronomical Society}, \textit{439}, 2515–2530. \href{https://arxiv.org/abs/1310.2820}{arXiv:1310.2820}

\bibitem{d115} Wang, Y., et al. (2017). The clustering of galaxies in the completed SDSS-III Baryon Oscillation Spectroscopic Survey: tomographic BAO analysis of DR12 combined sample in configuration space. \textit{Monthly Notices of the Royal Astronomical Society}, \textit{469}(3), 3762–3774. \href{https://arxiv.org/abs/1607.03154}{arXiv:1607.03154}

\bibitem{d116} Chuang, C.-H., \& Wang, Y. (2013). Modeling the Anisotropic Two-Point Galaxy Correlation Function on Small Scales and Improved Measurements of H(z), DA(z), and $\beta$(z) from the Sloan Digital Sky Survey DR7 Luminous Red Galaxies. \textit{Monthly Notices of the Royal Astronomical Society}, \textit{435}, 255–262. \href{https://arxiv.org/abs/1209.0210}{arXiv:1209.0210}

\bibitem{d117} Anderson, L., et al. (2014). The clustering of galaxies in the SDSS-III Baryon Oscillation Spectroscopic Survey: baryon acoustic oscillations in the Data Releases 10 and 11 Galaxy samples. \textit{Monthly Notices of the Royal Astronomical Society}, \textit{441}(1), 24–62. \href{https://arxiv.org/abs/1312.4877}{arXiv:1312.4877}

\bibitem{d118} Zhao, G.-B., et al. (2019). The clustering of the SDSS-IV extended Baryon Oscillation Spectroscopic Survey DR14 quasar sample: a tomographic measurement of cosmic structure growth and expansion rate based on optimal redshift weights. \textit{Monthly Notices of the Royal Astronomical Society}, \textit{482}(3), 3497–3513. \href{https://arxiv.org/abs/1801.03043}{arXiv:1801.03043}

\bibitem{d119} Busca, N. G., et al. (2013). Baryon Acoustic Oscillations in the Ly-$\alpha$ forest of BOSS quasars. \textit{Astronomy \& Astrophysics}, \textit{552}, A96. \href{https://arxiv.org/abs/1211.2616}{arXiv:1211.2616}

\bibitem{d120} Font-Ribera, A., et al. (2014). Quasar-Lyman $\alpha$ Forest Cross-Correlation from BOSS DR11: Baryon Acoustic Oscillations. \textit{Journal of Cosmology and Astroparticle Physics}, \textit{2014}(05), 027. \href{https://arxiv.org/abs/1311.1767}{arXiv:1311.1767}

\bibitem{d121} du Mas des Bourboux, H., et al. (2017). Baryon acoustic oscillations from the complete SDSS-III Ly$\alpha$-quasar cross-correlation function at z = 2.4. \textit{Astronomy \& Astrophysics}, \textit{608}, A130. \href{https://arxiv.org/abs/1708.02225}{arXiv:1708.02225}

\bibitem{d5} Alam, S., et al. (2017). The clustering of galaxies in the completed SDSS-III Baryon Oscillation Spectroscopic Survey: cosmological analysis of the DR12 galaxy sample. \textit{Monthly Notices of the Royal Astronomical Society}, \textit{470}(3), 2617–2652. \href{https://arxiv.org/abs/1607.03155}{arXiv:1607.03155}

\bibitem{d10} Chuang, C.-H., et al. (2013). The clustering of galaxies in the SDSS-III Baryon Oscillation Spectroscopic Survey: single-probe measurements and the strong power of normalized growth rate on constraining dark energy. \textit{Monthly Notices of the Royal Astronomical Society}, \textit{433}, 3559. \href{https://arxiv.org/abs/1303.4486}{arXiv:1303.4486}

\bibitem{d74} Blake, C., et al. (2012). The WiggleZ Dark Energy Survey: Joint measurements of the expansion and growth history at $z < 1$. \textit{Monthly Notices of the Royal Astronomical Society}, \textit{425}, 405–414. \href{https://arxiv.org/abs/1204.3674}{arXiv:1204.3674}

\bibitem{d79} Neveux, R., et al. (2020). The completed SDSS-IV extended Baryon Oscillation Spectroscopic Survey: BAO and RSD measurements from the anisotropic power spectrum of the quasar sample between redshift 0.8 and 2.2. \textit{Monthly Notices of the Royal Astronomical Society}, \textit{499}(1), 210–229. \href{https://arxiv.org/abs/2007.08999}{arXiv:2007.08999}

\bibitem{x38} Jimenez, R., \& Loeb, A. (2002). Constraining Cosmological Parameters Based on Relative Galaxy Ages. The Astrophysical Journal, 573(1), 37–42. \href{https://doi.org/10.1086/340549}{https://doi.org/10.1086/340549}


\bibitem{refS66} Burnham, K. P., \& Anderson, D. R. (2004). Multimodel inference: Understanding AIC and BIC in model selection. \textit{Sociological Methods \& Research}, \textit{33}(2), 261--304. \href{https://doi.org/10.1177/0049124104268644}{https://doi.org/10.1177/0049124104268644}

\bibitem{refS68} Liddle, A. R. (2004). How many cosmological parameters? \textit{Monthly Notices of the Royal Astronomical Society}, \textit{351}(3), L49--L53. \href{https://arxiv.org/abs/astro-ph/0401198}{astro-ph/0401198}

\bibitem{refS64} Schwarz, G. (1978). Estimating the dimension of a model. \textit{The Annals of Statistics}, \textit{6}(2), 461--464. \href{https://doi.org/10.1214/aos/1176344136}{https://doi.org/10.1214/aos/1176344136}


\bibitem{cc54} Zhang, C., Zhang, H., Yuan, S., Liu, S., Zhang, T.-J., Sun, Y.: Four new observational $H(z)$ data from luminous red galaxies in the Sloan Digital Sky Survey data release seven. \textit{Res. Astron. Astrophys.} \textbf{14}, 1221 (2014). \url{https://doi.org/10.1088/1674-4527/14/10/002}

\bibitem{cc55} Simon, J., Verde, L., Jimenez, R.: Constraints on the redshift dependence of the dark energy potential. \textit{Phys. Rev. D} \textbf{71}, 123001 (2005). \url{https://doi.org/10.1103/PhysRevD.71.123001}

\bibitem{cc56} Moresco, M., Cimatti, A., Jimenez, R., Pozzetti, L., Zamorani, G., Bolzonella, M., Dunlop, J.C., Lamareille, F., Mignoli, M., Pearce, H., Rosati, P., Stern, D., Verde, L., Zucca, E., Carollo, C.M., Contini, T., Kneib, J.-P., Le Fèvre, O., Lilly, S.J., Mainieri, V.: Improved constraints on the expansion rate of the Universe up to $z \sim 1.1$ from the spectroscopic evolution of cosmic chronometers. \textit{JCAP} \textbf{2012}, 006 (2012). \url{https://doi.org/10.1088/1475-7516/2012/08/006}

\bibitem{cc57} Alam, S., Ata, M., Bailey, S.J., Beutler, F., Bizyaev, D., Blazek, J., Bolton, A.S., Brownstein, J.R., Burden, A., Chuang, C.-H., Comparat, J., Cuesta, A.J., Dawson, K.S., Eisenstein, D.J., Escoffier, S., Gil-Marín, H., Grieb, J.N., Hand, N., Ho, S., Kinemuchi, K. et al.: The clustering of galaxies in the completed SDSS-III Baryon Oscillation Spectroscopic Survey: cosmological analysis of the DR12 galaxy sample. \textit{Mon. Not. R. Astron. Soc.} \textbf{470}, 2617 (2017). \url{https://doi.org/10.1093/mnras/stx721}

\bibitem{cc58} Moresco, M., Pozzetti, L., Cimatti, A., Jimenez, R., Maraston, C., Verde, L., Thomas, D., Citro, A., Tojeiro, R., Wilkinson, D.: A 6\% measurement of the Hubble parameter at $z \sim 0.45$: direct evidence of the epoch of cosmic re-acceleration. \textit{JCAP} \textbf{2016}, 014 (2016). \url{https://doi.org/10.1088/1475-7516/2016/05/014}

\bibitem{cc59} Ratsimbazafy, A.L., Loubser, S.I., Crawford, S.M., Cress, C.M., Bassett, B.A., Nichol, R.C., Väisänen, P.: Age-dating luminous red galaxies observed with the Southern African Large Telescope. \textit{Mon. Not. R. Astron. Soc.} \textbf{467}, 3239 (2017). \url{https://doi.org/10.1093/mnras/stx301}

\bibitem{cc60} Moresco, M.: Raising the bar: new constraints on the Hubble parameter with cosmic chronometers at $z \sim 2$. \textit{Mon. Not. R. Astron. Soc.} \textbf{450}, L16 (2015). \url{https://doi.org/10.1093/mnrasl/slv037}

\bibitem{planck2018} Planck Collaboration, Aghanim, N., Akrami, Y., et al. (2020). Planck 2018 results. VI. Cosmological parameters. \textit{Astronomy \& Astrophysics}, \textit{641}, A6. \href{https://arxiv.org/abs/1807.06209}{arXiv:1807.06209}

\bibitem{verde2019tensions} Verde, L., Treu, T., \& Riess, A. G. (2019). Tensions between the early and late Universe. \textit{Nature Astronomy}, \textit{3}, 891–895. \href{https://arxiv.org/abs/1907.10625}{arXiv:1907.10625}

\bibitem{farooq2013} Farooq, O., \& Ratra, B. (2013). Hubble parameter measurement constraints on the cosmological deceleration–acceleration transition redshift. \textit{The Astrophysical Journal Letters}, \textit{766}(1), L7. \href{https://arxiv.org/abs/1301.5243}{arXiv:1301.5243}

\bibitem{xu2012} Zheng, W., Li, H., Xia, J.-Q., Wan, Y.-P., Li, S.-Y., \& Li, M. (2014). Constraints on cosmological models from Hubble parameters measurements. \textit{International Journal of Modern Physics D}, \textbf{23}(05), 1450051. \href{https://doi.org/10.1142/s0218271814500515}{https://doi.org/10.1142/s0218271814500515}.

\bibitem{xu2012new} Kumar, D., Jain, D., Mahajan, S., Mukherjee, A., \& Rana, A. (2023). Constraints on the transition redshift using Hubble phase space portrait. \textit{International Journal of Modern Physics D}, \textbf{23}(06). \href{https://doi.org/10.1142/s0218271823500396}{https://doi.org/10.1142/s0218271823500396}

\bibitem{ref90AN} M. Koussour and N. Myrzakulov, ``Bouncing cosmologies and stability analysis in symmetric teleparallel $f(Q)$ gravity,'' \textit{Eur. Phys. J. Plus} \textbf{139}, 799 (2024). \href{https://doi.org/10.1140/epjp/s13360-024-05574-5}{https://doi.org/10.1140/epjp/s13360-024-05574-5}.

\bibitem{ref90AN1} L.~Lazkoz, F.~S.~N.~Lobo, M.~Ortiz-Baños, and V.~Salzano,  
``Observational constraints of $f(Q)$ gravity,''  
\textit{Phys. Rev. D} \textbf{100}, 104027 (2019).  
\href{https://doi.org/10.1103/PhysRevD.100.104027}{https://doi.org/10.1103/PhysRevD.100.104027}.
‌ 






















\end{thebibliography}
\end{document}